\documentclass[conference]{IEEEtran}
\IEEEoverridecommandlockouts
\usepackage{cite}
\usepackage{enumitem}
\usepackage{amsmath,amssymb,amsfonts,amsthm}
\usepackage{algorithm}
\usepackage{algorithmic}
\usepackage{graphicx}
\usepackage{textcomp}
\usepackage{xcolor}
\usepackage{booktabs}
\usepackage{subfigure}
\usepackage{float}
\usepackage{multirow}
\usepackage{url}
\usepackage{bm}
\usepackage{array}
\usepackage{graphicx}
\usepackage{nicefrac}
\usepackage{makecell}

\def\BibTeX{{\rm B\kern-.05em{\sc i\kern-.025em b}\kern-.08em
    T\kern-.1667em\lower.7ex\hbox{E}\kern-.125emX}}

\newcolumntype{L}[1]{>{\raggedright\let\newline\\\arraybackslash\hspace{0pt}}m{#1}}
\newcolumntype{C}[1]{>{\centering\let\newline  \\\arraybackslash\hspace{0pt}}m{#1}}
\newcolumntype{R}[1]{>{\raggedleft\let\newline \\\arraybackslash\hspace{0pt}}m{#1}}

\newtheorem{prop}{Proposition}

\newtheorem{definition}{Definition}

\usepackage[normalem]{ulem}

\begin{document}

\title{Knowledge-Enhanced Recommendation with 
	User-Centric Subgraph Network}

\author{\IEEEauthorblockN{Guangyi Liu\IEEEauthorrefmark{2},
		Quanming Yao\IEEEauthorrefmark{3}, Yongqi Zhang\IEEEauthorrefmark{4} and
		Lei Chen\IEEEauthorrefmark{5}}
	
	\IEEEauthorblockA{
		\IEEEauthorrefmark{2}\IEEEauthorrefmark{3}Department of Electronic Engineering, Tsinghua University, Beijing, China\\
		\IEEEauthorrefmark{4}4Paradigm Inc., Beijing, China\\
		\IEEEauthorrefmark{5}The Hong Kong University of Science and Technology (GZ), Guangzhou, China\\
		\{\IEEEauthorrefmark{2}liugy20@mails. \IEEEauthorrefmark{3}qyaoaa@\}tsinghua.edu.cn, 
		\{\IEEEauthorrefmark{4}yzhangee@connect. \IEEEauthorrefmark{5}leichen@\}ust.hk
	}
}
\maketitle

\begin{abstract}
 
Recommendation systems, as widely implemented nowadays on
various platforms, recommend relevant items to users based on
their preferences.
The \textcolor{black}{classical} methods which rely on user-item interaction matrices
	 has limitations, 
	especially in scenarios where there is a lack of interaction data for new items.
Knowledge graph (KG)-based recommendation systems have emerged as a promising solution. 
However, most KG-based methods adopt node embeddings, 
which do not provide personalized recommendations for different users 
and cannot generalize well to the new items.
To address these limitations, 
we propose Knowledge-enhanced User-Centric subgraph Network (KUCNet),
\textcolor{black}{a subgraph learning approach with graph neural network (GNN)
for effective recommendation.} 
KUCNet constructs a U-I subgraph for each user-item pair that captures both the historical information of user-item interactions and the side information provided in KG. 
An attention-based GNN is designed to encode the U-I subgraphs
for recommendation.
Considering efficiency,
the pruned user-centric computation graph
is further introduced such that  multiple U-I subgraphs can be simultaneously computed
and that the size can be pruned by Personalized PageRank.
Our proposed method achieves accurate, efficient, and interpretable recommendations
especially for new items. 
Experimental results demonstrate the superiority of KUCNet over state-of-the-art KG-based and collaborative filtering (CF)-based methods.
Our code and data is available in \url{https://github.com/leolouis14/KUCNet}.
\footnote{This work was partially done when G. Liu was an intern
	at 4Paradigm, and the correspondence is to Y. Zhang.}

\end{abstract}


\section{Introduction}

\begin{figure}[t]
	\centering
	\includegraphics[width=1\linewidth]{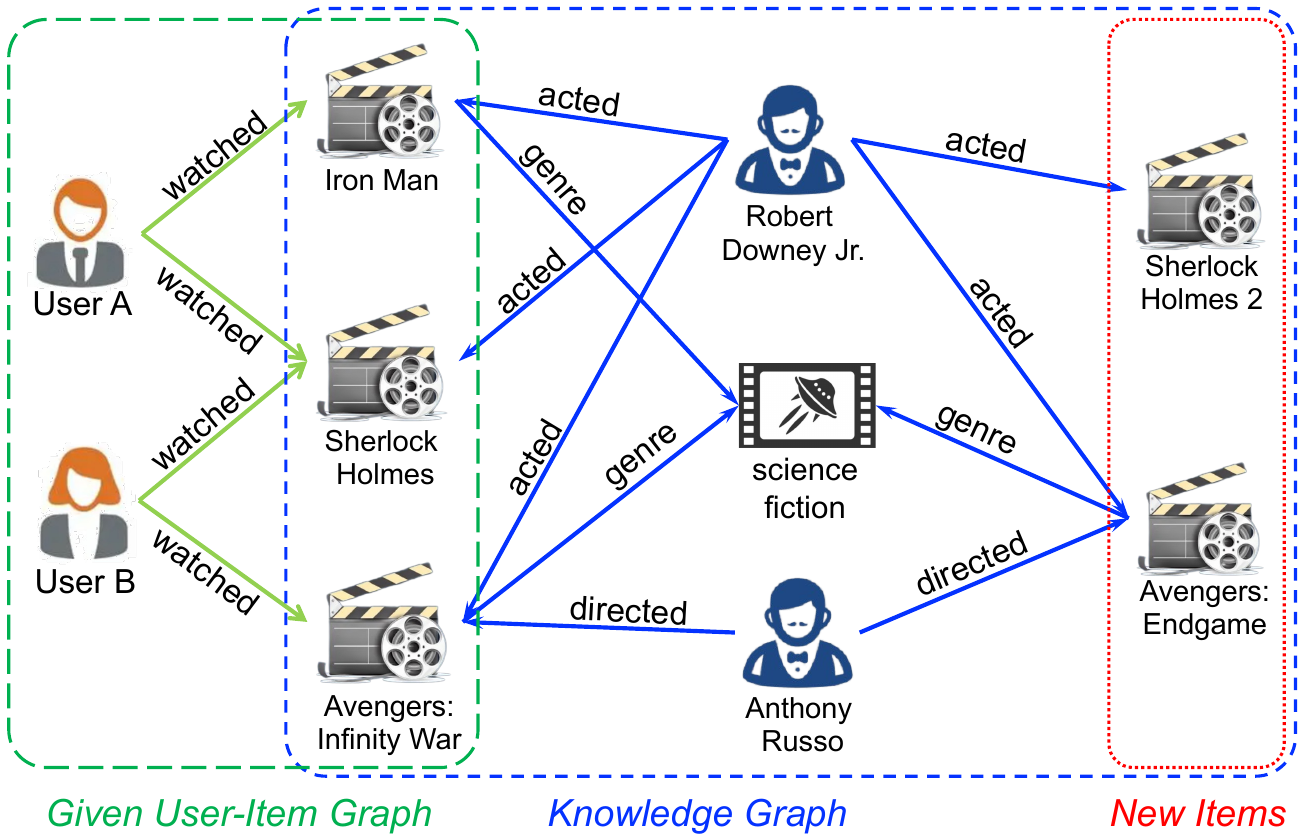}
	\vspace{-8px}
	\caption{Illustration of knowledge graph enhanced recommender systems.
		In the user-item graph (in green), the interactions between two users and three items are provided.
		The knowledge graph (in blue) provides the interacted entities as side information for the items.
		In particular,
		there are a few new items (e.g. newly released movies) that not appear in the user-item graph but can be connected with given items with the entities in KG.
		The task is to recommend items to \textit{UserA} or \textit{UserB}.}
	\label{fig:example}
	\vspace{-4px}
\end{figure}

Recommendation systems have become an essential part of our daily lives, 
suggesting everything from movies to watch \cite{levandoski2012lars}, 
products to buy \cite{wang2022tower}, 
to even friends to connect with on social media \cite{hui2022time}. 
These systems are designed to analyze users' historical behavior and preferences, 
and provide personalized recommendations to enhance user experience \cite{resnick1997recommender,isinkaye2015recommendation}.
With the exponential growth of data, the need for 
accurate recommendation systems has increased tremendously. 
The classical matrix factorization based methods
 \cite{he2017neural,he2018nais} 
has been widely used in recommendation systems 
and has achieved significant success. 
This approach relies on users' historical interactions 
and assumes that users who have similar behaviors in the past 
are likely to have similar interests in the future \cite{koren2009matrix,rendle2012bpr,rendle2011fast,cao2018neural}.  
However,
the performance of these methods is not as good as desired since the user-item interaction is sparse, 
particularly in scenarios where there is a lack of interaction data for 
new items
like the example in Figure~\ref{fig:example}.
In this example, the new items \textit{Sherlock Holmes2} and \textit{Avengers: Endgame} have no given interaction with the users.
In addition,
these collaborative filtering (CF)-based methods are hard to interpret from the high-dimensional representations of users and items.

In recent years, 
knowledge graph (KG)-based recommendation systems have emerged as a promising solution 
to address the limitations of traditional recommendation systems \cite{zhang2016collaborative,guo2020survey}. 
KG is a type of directed multi-relational graph 
that contains a wealth of rich facts 
and can provide ample side information to enhance the interactions between users and items~\cite{wang2017knowledge,ji2021survey}.
By incorporating KG information, 
recommendation systems can improve accuracy, diversity, and explainability of recommendations, 
and alleviate the problems of data sparsity
and new instances.
For example, in the KG shown in Figure~\ref{fig:example}, 
the new items \textit{Sherlock Holmes 2} and \textit{Avengers: Endgame} have no given interaction with the users. 
However, the KG provides information for these items, 
offering connections between them and items in the given user-item graph. 
Therefore, 
the rich semantic information and structural data on the KG can help overcome the limitations of traditional recommendation systems 
and provide more personalized and relevant recommendations to users.

Earlier KG-based recommendation methods, 
such as CKE \cite{zhang2016collaborative}, 
used KG embedding techniques to measure the plausibility of interactions and triplets in KG with embeddings. 
However, 
these methods are shallow as they only work on individual edges. 
More recently, 
several methods \cite{wang2019kgat,wang2019knowledge,wang2020ckan,wang2021learning} 
have leveraged graph neural networks (GNNs)\cite{kipf2016GNN1,hamilton2017GNN2,he2020GNN3}
to build the recommender model.
GNNs in these methods aggregate information from neighboring nodes
to update the representation of users and items, 
allowing the side information in KG to be incorporated into the representations. 
However, most of the GNN-based methods adopt node embedding, 
which does not vary with users,
and also employ global aggregation, which fails
to capture the intricate relation structures between users and
candidate items, therefore does not provide personalized recommendations for different users.
And the node embedding limits their performance in the scenario where there are new items, 
as the lack of supervision signals makes it hard to learn the representation of new items accurately.
Additionally, 
performing global aggregation may introduce noisy information in KG that is not relevant for recommendation  \cite{wang2019kgat,ji2021survey}, 
which can negatively affect the accuracy 
and effectiveness of the recommendation system.

To better leverage both the historical information 
of user-item interactions and the side information in KG, 
and address the limitations mentioned above,
we propose a personalized representation learning method for each user-item pair to capture the most relevant information. 
Firstly,
we define a U-I subgraph for each user-item pair, 
which contains both 
the collaborative similarity existing in user-item interaction
and the attribute similarity existing in the KG.
In comparison,
most of the node embedding aggregation methods
do not consider both parts.
However, 
the main challenge is computation efficiency, 
as there can be many candidate items to evaluate, 
and the subgraph size may be large. 
To solve this problem, 
we propose the Knowledge-enhanced User-Centric subgraph Network
(KUCNet)
to achieve efficient and interpretable subgraph learning.
Specifically,
we introduce U-I subgraph for each user-item pair
and design an attention-based GNN to encode the subgraphs for recommendation.
To improve efficiency,
we introduce a pruned user-centric computation graph 
that allows simultaneous encoding of multiple U-I subgraphs and 
use Personalized PageRank to reduce the graph size. 
In this way, 
the proposed method achieves accurate, efficient, 
and interpretable recommendation 
with pairwise subgraph encoding.
{\color{black}
Moreover, not using node embedding 
allows KUCNet to be generalized to two special 
recommendation scenarios: recommendation with new items and new users,
which have practical significance but have been consistently overlooked by KG-based recommendation methods.
Experiments shows that existing methods cannot work properly in these two scenarios, 
but our method can perform excellently.}
The contributions are summarized as follows:
\begin{itemize}
{\color{black}
	\item 
	We propose a 
	subgraph learning approach 
	with a novel data structure 
	called the U-I subgraph 
	for knowledge-enhanced recommendation. 
	The U-I subgraph preserves 
	personalized information from
	both the related user-item interactions and KG side-information.

	\item \textcolor{black}{We design a user-centric subgraph network that allows 
	efficient encoding on multiple pruned U-I subgraphs 
	and weights the importance of edges in the subgraphs. 
	Consequently, KUCNet can achieve accurate, efficient, and interpretable recommendations.}

	\item 
	Experimental results demonstrate that KUCNet 
	is effective in recommendation, 
	outperforming state-of-the-art KG-based 
	and CF-based recommendation methods,
	while maintaining high efficiency and interpretability.
	
	\item
	Apart from the traditional recommendation,
	we set up scenarios for recommending with new items and users.
	The results demonstrate the importance of KG
	and the superiority of KUCNet in recommending with new items and users.
}
\end{itemize}

\vspace{-5px}
\section{Related work}
\label{sec:relwork}

\subsection{Traditional methods for recommendation}
Recommendation system is designed to provide suggestions
of items to users
depending on previous interaction.
Collaborative filtering (CF)\cite{he2017neural,he2018nais} is a popular approach in recommendation systems that 
relies on analyzing the historical behavior of users/items to identify the similarities 
between them. 
For example, 
Matrix Factorization (MF)~\cite{koren2009matrix,rendle2012bpr} is a widely-used 
collaborative filtering method that decomposes the user-item interaction matrix into 
low-rank matrices to capture users' and items' features. 
Another traditional CF-based method, 
Factorization Machines (FM)\cite{rendle2011fast} 
considers contextual information to provide context-aware rating predictions.
And it can model higher-order feature interactions, which helps it remain effective 
in handling high-dimensional and sparse data. 
Neural Factorization Machines (NFM)\cite{cao2018neural} is an extension of FM by 
leveraging neural networks to learn more complex feature interactions and further 
improve model performance and generalization.
Since these CF-based methods are effective to capture the user preference 
and can be easily implemented, 
they have been widely used in many scenarios.
However,
these methods rely heavily on the historical interactions between users and items, and implicitely encode these interactions
into latent representations (a.k.a. embeddings) of users/items.

PageRank algorithm~\cite{page1997pagerank} is a link analysis algorithm 
widely used in web search engines and other applications. 
It evaluates each webpage 
by random walk and iterative calculation, assigning higher scores to more important webpages 
and then sorting search results accordingly.
Personalized PageRank (PPR)~\cite{haveliwala200ppr} is a variant of the PageRank algorithm that 
aims to compute personalized PageRank scores for each specific target node. 
The personalized scores can indicate the amount of importance other nodes
to the target node.
In recommendation systems,
some works~\cite{baluja2008video,bogers2010movie,li2013recommendation,musto2021PPR}
utilize it to calculate importance scores of items 
related to the interests of each user, 
which can be used to recommend items for users. 
But this approach is heuristic
and may not be able to discover latent properties or correlations
within users and items.

\subsection{Knowledge graphs (KG) for recommendation}

KG is a kind of semantic network representing
relationships between real-world entities,
thus can also represent the interactions for users and items,
and provide semantic relatedness among items.
With the help of KG,
the latent connections and relevance of items to users
can be more abundant.
Hence, this inspires many recent works
to leverage KG for recommendation systems
and propose new approaches to learn from KGs.
By leveraging KG embedding techniques,
CKE\cite{zhang2016collaborative} 
exploits semantic embeddings derived from TransR\cite{lin2015TransR} for recommendation.
This method only leverages the first-order connectivity
(i.e., single edges with user-item pairs in CF or 
triplets in KG),
thus is shallow and not very expressive.
Rather than using direct connections,
MCRec\cite{hu2018leveraging} 
extracts some pre-defined patterns of paths (a.k.a. meta-paths) as features and
utilizes 
a convolutional layer to encode the features into interactions.
RippleNet~\cite{wang2018ripplenet}
and 
TB-Net\cite{wang2022tower}
 learns embeddings
and propagate along the paths in KG,
which automatically discovers users’ hierarchical 
potential interests 
and enrich user representations.
However,
these methods are expensive in paths finding
and modeling,
and do not well capture the helpful structural patterns in KG.

With the development of GNNs~\cite{kipf2016GNN1,hamilton2017GNN2,he2020GNN3} 
in modeling graph structured data,
several methods ~\cite{wang2019knowledge, wang2019kgat,wang2020ckan,wang2021learning}
ultilize GNN to aggregate information 
for users and items from KG.
Typically, they incorporate information from neighboring 
nodes to update the representation of a self node. 
In this way, information from 
distant nodes can be encoded in the representation. 
Therefore, these methods
like KGNN-LS~\cite{wang2019knowledge}
and CKAN~\cite{wang2020ckan}
are able to model remote connections.
And KGAT\cite{wang2019kgat} uses an attention mechanism with KG relations to  discriminate the importance of the
neighbors, then propagates the embeddings on the holistic graph. 
KGIN~\cite{wang2021learning} models user’s different intents 
and performs the path-aware aggregation for both user-item graph and KG.
The attention mechanism brings higher effectiveness and better interpretability.
Since all these methods adopt node embedding 
for each user and item, they do not well provide 
personalized recommendations
for different users.
In addition,
they cannot work on the cases where new users or new items may appear.

\subsection{Link prediction in KG}
\label{ssec:relatedKG}

Link prediction is a common task in KG
whose objective is to predict missing link from an incomplete KG.
Recommendation can be regarded as a link prediction problem
targeting on the interaction edges between users and items.
The earlier methods for link prediction
are embedding methods,
which map entities and relations to low-dimensional representation vectors 
and design a scoring function to measure the plausibility of triplets (single edges).
Representative methods
like TransE~\cite{bordes2013transE}, TransR~\cite{lin2015TransR}
design different kinds of scoring function to measure such plausibility in different ways.
These methods are hard to interpret due to the high-dimensional embedding space.
Besides, their generalization ability to inductive case is low 
since specific embeddings are required for each entity.

The recent trend in link prediction turns from scoring function design on triplets to graph learning.
R-GCN \cite{schlichtkrull2018RGCN}
and CompGCN \cite{vashishth2019composition}
apply GNNs to aggregate information on KGs,
but do not show superior advantage over the embedding methods.
Based on the subgraph learning method SEAL~\cite{zhang2018link},
GraIL~\cite{teru2020inductive} was proposed to leverage subgraphs in KG to measure the plausibility of target triplets.
REDGNN~\cite{zhang2022knowledge} and NBFNet~\cite{zhu2021neural} improve the efficiency of GraIL
by dynamic programming.
These methods show that subgraphs can be more effective than
embeddings.
In addition,
the subgraphs are inductive since the prediction with subgraph
does not need to learn specific embeddings for each entity.

\begin{figure*}[t]
	\centering
	\includegraphics[height=5.3cm]{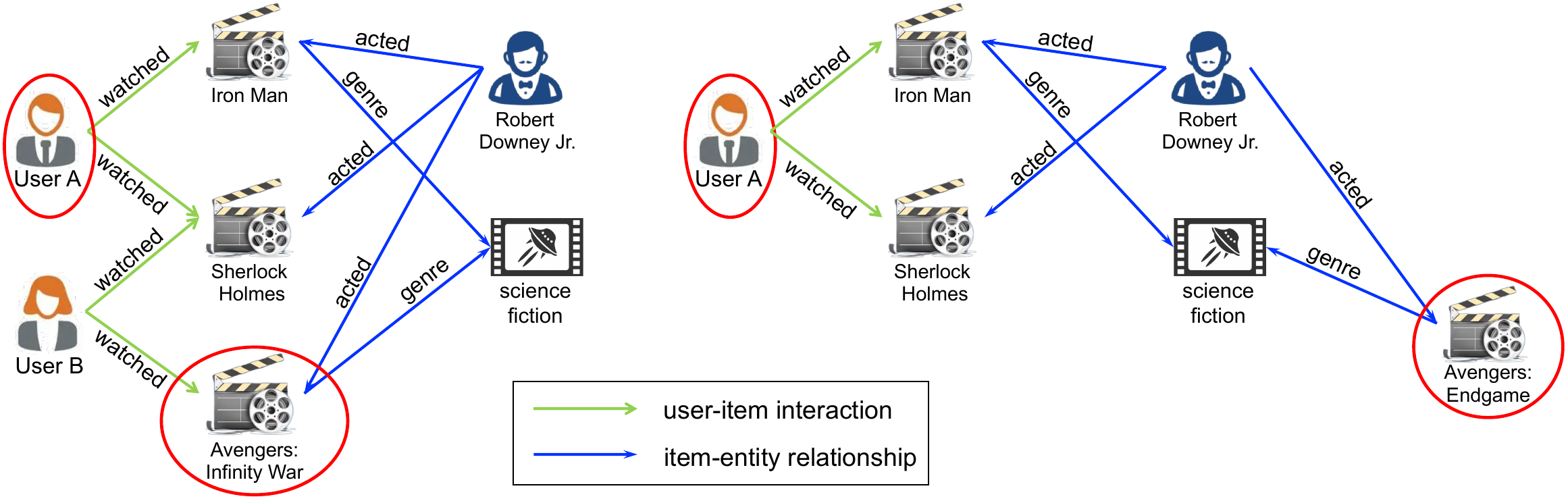}
	\vspace{-6px}
	\caption{
		Examples of U-I subgraph for \textit{UserA} with two different items.
		[Left]: This subgraph contains both the collaborative filtering information about the common interests between user A and user B, 
		as well as the similarity among the movies in terms of actor and genre.
		[Right]: This subgraph provides the connection between an new item with given items that \textit{UserA} watched with knowledge in KG.}
	\label{fig:UIgraph}
	\vspace{-7px}
\end{figure*}

\section{Problem formulation}

We first introduce three structural data: user-item graph, knowledge graph and
collaborative knowledge graph, 
and then formulate the knowledge-enhanced recommendation task. 
Figure~\ref{fig:example} provides an example of the problem studied in this paper.

\textit{User-Item Graph.} 
In a recommendation task, we have historical user-item interactions
under the  implicit feedback setting\cite{rendle2012bpr},
e.g., purchase, view,
clicks.
That is, we have user set $\mathcal U$, item set $\mathcal I$
and a set of observed feedback 
$\mathcal O^+=\{(u,i)|u \in \mathcal U, i \in \mathcal I\}$, 
where each $(u,i)$ pair indicates that
user $u$ has interacted with item $i$.
Here we represent interaction data as a user-item bipartite
graph $\mathcal E_\text{ui}$ with relation ``interact'', 
which is defined as 
$\mathcal E_\text{ui} = \{(u, interact,i)|(u,i)\in \mathcal O^+\}$.
\textcolor{black}{
As such, the user-item interactions can be seamlessly combined with KG
to communicate with the heterogeneity information.}

{\color{black}
\textit{Knowledge Graph (KG)}. In addition to the user-item interactions, we have side
information for items (e.g., attributes and external relevant knowledge of items).
For example, a movie can be described by entities of its director, actors and genres.
This information is organized into a Knowledge Graph $\mathcal G_k = (\mathcal V_k, \mathcal E_k)$,
where $\mathcal V_k$ is a set of real-world entities, and
$\mathcal E_k = \{(h, r, t)|h, t \in \mathcal V_k, r \in \mathcal R_k\}$ is a collection of triplets,
where $\mathcal R_k$ is the relation set,
and each $(h, r, t)$ describes that there is a relationship $r$ from head entity $h$ to
tail entity $t$.

\textit{Collaborative Knowledge Graph (CKG)}. 
Given User-Item Graph and Knowledge Graph, 
we integrate them to construct a Collaborative Knowledge Graph (CKG) $\mathcal G$.
We follow prior work\cite{wang2021learning} to establish a set of item-entity matching 
$\mathcal{M} = \{(i, e)|i \in \mathcal{I}, e \in \mathcal V_k\}$, 
where $(i, e)$ indicates that item $i$ can be aligned with an entity $e$ in the KG.
In this way, two different graphs can be connected to form a collaborative graph.
Note that users and certain items may not have corresponding entities, 
but they can still be integrated into CKG through those mapped items.
Let $\mathcal V$ be a set of nodes ($\mathcal V = \mathcal U \cup \mathcal I \cup \mathcal V_k$)
and $\mathcal R$ be a set of relations ($\mathcal R=\mathcal R_k \cup \text{``interact''}$).
We define $\mathcal E = \{(n_s,r,n_o)| n_s,n_o \in \mathcal V, r\in\mathcal R \}$ as a set of triplets (edges) in the CKG 
$\mathcal G = (\mathcal V, \mathcal E)$.
It is obvious that
$\mathcal E_\text{ui}\subset \mathcal E$.
}
\begin{definition}[Knowledge-enhanced recommendation]
\label{def:KGrec}
Given 
the CKG $\mathcal G$
including the interaction graph $\mathcal E_\text{ui}$, 
our task is to learn a prediction function $\mathcal F$
that 
outputs the probability 
\begin{equation}
\hat{y}_{ui} = \mathcal{F}(u,i|\mathcal G),
\end{equation}
that user $u$ would adopt item $i$.
\end{definition}

CKG contains rich semantic and structural information that can 
enhance the interactions between users and items
and improve the accuracy of recommendations,
especially for the new items that have no interaction with users.
However, 
the CKG can be large and noisy \cite{wang2019kgat,ji2021survey}, 
and deciding which information to use is crucial.
Existing GNN-based methods that adopt node embedding
do not consider the personality of information in CKG for recommendation
and
aggregate all the side information.
In addition,
few of them  
consider both the user-item interaction as well as KG's side information
simultaneously.
In the next part,
we will propose a new method based on subgraph learning
to address these limitations.

{\color{black}
In this paper,
we denote vectors by lowercase boldface,
e.g. $\bm h, \bm w, \bm r$,
and matrix by upper case boldface,
e.g. $\bm W, \bm M$.
The other commonly used notations
are summarized in Table~\ref{tab:notation}.
}
\begin{table}[t]
	\centering
	\caption{Frequently used notations in this paper.}
	\label{tab:notation}
	\vspace{-8px}
	\renewcommand\arraystretch{1.1}
	\begin{tabular}{c|l}
		\toprule
		Symbols  & Meanings\\
		\midrule
		$\mathcal U, \mathcal I$ & the set of users/items \\
		$\mathcal V, \mathcal R,\mathcal E$ & the set of entities/relations/edges in collaborative KG \\
		$(u, i)$ & a pair of interaction between user $u$ and item $i$ \\
		$y_{ui}, \hat{y}_{ui}$ & true/predicted label between $u$ and $i$ \\
		$(n_s, r, n_o)$ &  an edge in the collaborative KG \\
		$\mathcal G_{u,i|L}$
		& an $L$-layer U-I subgraph of $(u,i)$\\
		$\bm h_{u:o}^l$ & the representation propagated from $u$
		to node $o$ in layer $l$\\
		${\mathcal C}_{u,i|L}$
		& the computation graph on $\mathcal G_{u,i|L}$ \\
		${\mathcal{C}}_{u|L}$ &
		the user-centric computation graph of user $u$ with $L$-layer  \\
		$\widetilde{\mathcal{C}}_{u|L}$ &
		the pruned user-centric computation graph of ${\mathcal{C}}_{u|L}$ \\
		\bottomrule
	\end{tabular}
	\vspace{-6px}
\end{table}

\section{Proposed method}

We now present the proposed Knowledge-enhanced User-Centric subgraph Network
(KUCNet).
We first introduce the concept of U-I subgraph in Section \ref{ssec:original-uisubgraph}, 
which contains both the collaborative signals and external knowledge.
We then customize the GNN architecture with an attention mechanism
to encode subgraphs in Section \ref{ssec:gnn}.
To improve computation efficiency,
Section \ref{ssec:uisubgraph}
shows how we construct a pruned user-centric subgraph, 
that allows simultaneous computation of multiple U-I subgraphs and is truncated by Personalized PageRank. 
Finally, we show the details of algorithm and optimization
in Section \ref{ssec:optimize}.

\subsection{Subgraph learning for recommendation}
\label{ssec:original-uisubgraph}

Existing GNN-based methods for recommendation
mainly focus on aggregating the related entities 
in KG for items as supplementary information.
While prior works\cite{wang2019kgat,wang2020ckan,wang2021learning} learn to weight the different edges
in KG with attention mechanism,
these methods do not fully consider the 
pairwise influence between users and items.
For example, 
considering the case 
where a user likes science fiction movies
and another user prefers comedies,
the same science fiction movie or comedy
will have distinct meanings for the two users.
Hence,
it is crucial to
model relationships separately with fine-grained granularity 
for each target user and candidate item.

To achieve this
we can build a specific data structure, 
like the subgraph discussed in Section~\ref{ssec:relatedKG}
that preserving essential knowledge related to the target 
user and item,
for recommendation.
Inspired by the success of GNN-based subgraph learning~\cite{zhang2018link,teru2020inductive,zhang2022knowledge},
we introduce user-item subgraph for given user, 
and item
and customize the GNN architecture for 
interpretable and capable  recommendation on new items
by learning from the user-item subgraphs.
\begin{definition}[U-I subgraph]
Given a pair $(u,i)$ of user-item,
we define 
\begin{equation}
\mathcal G_{u,i|L} = \left(\mathcal V_{u,i|L}, \mathcal E_{u,i|L} \right),
\end{equation}
as the user-item subgraph between $u$ and $i$ with maximum
depth $L$.
Here $\mathcal V_{u,i|L}\subset\mathcal V$ contains all the nodes whose 
sum of shortest-path distance to both $u$ and $i$
is no larger than $L$,
and $\mathcal E_{u,i|L}\subset \mathcal E$ contains all the
edges connecting nodes in $\mathcal V_{u,i|L}$.
\end{definition}

Figure~\ref{fig:UIgraph} provides examples of the U-I subgraphs
extracted from Figure~\ref{fig:example}.
As shown,
both the subgraphs contain essential information
for recommendation
based on either similar user behaviors or similar entity relationships.
Specifically,
we can get access to two kinds of typical side information in the subgraph:
\begin{itemize}[leftmargin=*]
	\item one is collaborative similarity, e.g., both \textit{UserA} and \textit{UserB} like  \textit{Sherlock Holmes};
	\item the other one is attribute similarity, e.g., 
	both \textit{Iron Man} and \textit{Avengers: Endgame} belong to the genre \textit{science fiction}.
\end{itemize}
The two types of information are often
overlooked or not simultaneously considered by
the existing KG-based recommender methods \cite{wang2019knowledge,wang2020ckan,wang2021learning}.
With the defined U-I subgraph,
our problem is specified as learning a
graph classifier (GNN to be exact)
$\mathcal F(\cdot)$ with parameters $\bm \Theta$
that maps the subgraph $\mathcal G_{u,i|L}$
into the interaction probability of $u$ and $i$
\begin{equation}
\hat{y}_{ui} = \mathcal F(\mathcal G_{u,i|L}; \bm \Theta).
\label{eq:graph}
\end{equation}

Given a pair of user-item $(u,i)$,
Eq.\eqref{eq:graph} is composed of the following steps:
(i) construct the subgraph $\mathcal G_{u,i|L}$ from the CKG;
(ii) build a GNN classifier $\mathcal F(\cdot)$  with parameter 
$\bm \Theta$,
and run message passing for $L$ steps on $\mathcal G_{u,i|L}$.
Prior works  \cite{zhang2018link,teru2020inductive,zhang2022knowledge} on subgraph learning
are not well incorporated into recommendation systems.
This work is a first attempt to adopt the power of subgraph learning
for knowledge-enhanced recommendation.
Compared to the global aggregation methods \cite{wang2019knowledge,wang2019kgat,wang2021learning},
the U-I subgraph can provide more personalized and detailed information
for accurate recommendation,
predict without relying on specific user or item's embeddings
for recommending new users or items,
and can be interpretable as will be shown in Section~\ref{ssec:exp:interpret}.

\subsection{Message passing on the U-I subgraph}
\label{ssec:gnn}

To enable better communication 
among users, items and entities,
we follow  the common practices
in KG learning \cite{vashishth2019composition,zhang2022knowledge}
to introduce reverse relations (denoted as $-r$ for each $r\in\mathcal R$) in the CKG.
With the reverse edges,
a user $u$ can arrive at $i$ in exactly $L$ jumps,
e.g. through the paths
$u-i-u-i$ or $u-i-e-i$ when $L=3$.
Formally,
we denote the paths pointing from $u$ to $i$ with length $L$ in the form
\begin{equation}
	u\xrightarrow{r^1} n^1\xrightarrow{r^2} n^2\cdots n^{L-1}\xrightarrow{r^{L}} i,
	\label{eq:path}
\end{equation}
where the $l$-th edge is denoted as $e^l=(n^{l-1},r^{l},n^l)$
with $n^0=u$ and $n^L = i$.
Let $\mathcal{E}_{u,i|L}^l=\{e^l: e^l=(n^{l-1},r^{l},n^l)\}$ be the set of all edges at hop $l$ ($l=1,\cdots, L$) in the paths of Eq.\eqref{eq:path}.
Similarly, let $\mathcal{V}_{u,i|L}^l=\{n^l: (n^{l-1},r^{l},n^l) \in \mathcal{E}_{u,i|L}^l\}$ 
be the set of tail nodes at hop $l$.
Specially, we have $\mathcal{V}_{u,i|L}^0=\{u\}$ and $\mathcal{V}_{u,i|L}^L =\{i\}$.

In order to leverage both the collaborative information
and side information,
and learn specific representation for the given pair of $(u,i)$,
we propagate the representation from user $u$ to item $i$
while aggregating the intermediate information in 
{\small $\mathcal{E}_{u,i|L}^l, l=1,\dots L$}. 
Specifically,
we define the message passing mechanism \cite{kipf2016GNN1,Gilmer2017}
as
\begin{equation}
	\bm{h}^l_{u:o}=\delta\Big(\sum\nolimits_{(n_s,r,n_o)\in \mathcal{N}_{o}^l}\varphi\big(\bm{h}^{l-1}_{u:s},\bm{h}_r^l\big)\Big),
	\label{eq:mpnn}
\end{equation} 
where 
{\small $\mathcal{N}_{o}^l = \{(n_s,r,n_o)\in{\mathcal E}_{u,i|L}^l \}$}
is the set of neighboring edges of tail node $n_o$ in the edge set ${\mathcal E}_{u,i|L}^l$,
$\delta(\cdot)$ is the activation function, e.g., ReLU,
$\bm{h}_r^l$ is the learnable embedding of relation $r$ in the $l$-layer.
$\varphi(\cdot)$ is a message function computing 
the message of the neighboring edges 
in $\mathcal{N}_{o}^l$
based on the  nodes'
and relation's embeddings.
The representation $\bm{h}^l_{u:o}$ means the 
propagated representation from user $u$ to any node 
{\small $o\in\mathcal V_{u,i|L}^l$} at the $l$-th layer.
{\color{black} Note that all $\bm{h}^l_{u:o}$'s are not node embeddings 
	but relative node representations that are calculated based on the U-I subgraph.
In this way, our model can be generalized to scenarios 
where new items/users occurred, since we do not require embeddings
of new nodes. }

In recommendation,
it is crucial to interpret why the model derives such a recommending results.
{\color{black}In order to 
	differentiate the different roles of edges,}
an attention mechanism is used 
to provide different weights to the different
edges in the recommendation routine.
Specifically,
we define the message function $\varphi(\cdot)$ as 
\begin{equation}
	\varphi(\bm{h}^{l-1}_{u:s},\bm{h}_r^l)
	=\alpha_{u:sr}^l \bm{W}^l(\bm{h}^{l-1}_{u:s}+\bm{h}_r^l),
	\label{eq:message func}
\end{equation}
where $\bm{W}^l\in \mathbb{R}^{d\times d}$, and $\alpha_{u:sr}^l$ is attention weight on 
the different $(n_s,r,n_o)$ in $\mathcal{N}_{o}^l$. 
The attention weights are defined with a parameterized function,
such that the inherent importance could be captured, i.e.,
\begin{equation*}
	\alpha_{u:sr}^l=\sigma\left((\bm{w}_\alpha^l)^{T}\text{ReLU}(\bm{W}_{\alpha s}^l \bm{h}^{l-1}_{u:s}
	+\bm{W}_{\alpha r}^l \bm{h}^{l}_{r}+\bm{b}_{\alpha})\right),
\end{equation*}
where $\sigma$ is the sigmoid function 
weighting the edge with value in interval $[0,1]$, 
$\bm{w}_\alpha^l \in \mathbb{R}^{d_\alpha}$,
$\bm{W}_{\alpha s}^l, \bm{W}_{\alpha r}^l \in \mathbb{R}^{d_\alpha\times d}$ and $\bm{b}_\alpha\in \mathbb{R}^{d_\alpha}$
are the learnable weighting matrices or vectors.
In this way,
we can weight the importance of different edges
\textcolor{black}{to  distinguish the roles of different nodes and edges
in the subgraph.}
After $L$-steps propagation with Eq.\eqref{eq:mpnn},
we can obtain the pair-wise encoding
$\bm h_{u:i}^L$ as the  encoding of subgraph $\mathcal G_{u,i|L}$.
A linear layer then maps the representation $\bm{h}^L_{u:i}$
to the probability logit of interaction between user $u$
and item $i$, i.e.,
\begin{equation}
	\hat{y}_{ui} = \bm{w}^T \bm{h}^L_{u:i},
	\label{eq:pred}
\end{equation}
where the weighting vector $\bm w\in\mathbb R^d$.

\begin{figure*}[t]
	\centering
	\includegraphics[width=1\linewidth]{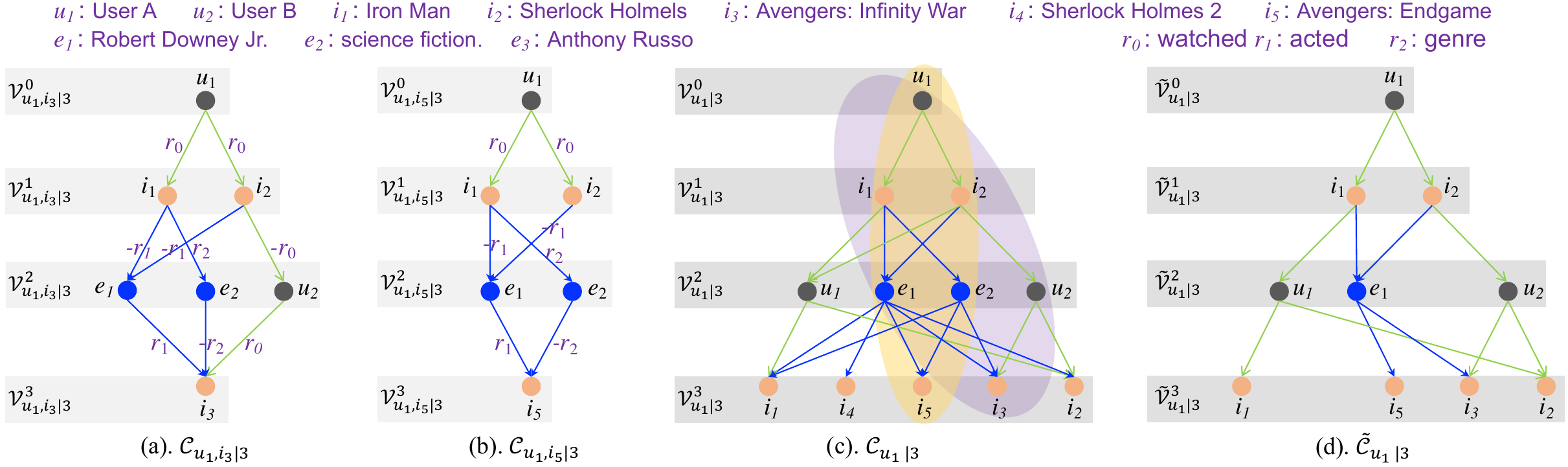}
	\vspace{-10px}
	\caption{ 
		Graphical illustration of computation graphs
		derived from Figure~\ref{fig:example}.
	(a). the computation graph on $\mathcal G_{u_1, i_3|3}$;
	(b). the computation graph on $\mathcal G_{u_1, i_5|3}$;
	(c). the user-centric computation graph with ellipses indicating
that ${\mathcal C}_{u_1,i_3|3}$ and ${\mathcal C}_{u_1,i_5|3}$
are subgraphs of $\mathcal C_{u_1|3}$;
	(d). the pruned user-centric computation graph with $K=2$.}
	\label{fig:pathgraph}
	\vspace{-6px}
\end{figure*}

\subsection{Improving the computation efficiency}
\label{ssec:uisubgraph}

When implementing the message passing in Section~\ref{ssec:gnn},
a big challenge we face is the 
efficiency problem.
Denote $\bar{C}$ as the average number of edges in 
different $\mathcal G_{u,i|L}$,
the complexity of recommending items for each user $u$
is $O(|\mathcal I|\bar{C})$
since such a process should be conducted for 
different items in $\mathcal I$.
The computation cost lies in two aspects:
the first is that the number of items $|\mathcal I|$ 
and the second is the size of subgraphs.

 \begin{itemize}[leftmargin=15px]
 	\item[(i)] \textbf{Multiple evaluation items}: As a recommendation system, the general objective is to recommend the most potential items from the item set $\mathcal I$.
 	Hence, it will be expensive to solve Eq.\eqref{eq:graph} when many items
 	need to be evaluated.
 	\item[(ii)]  \textbf{Subgraph size}: The subgraph size might be large. If the user interacted with some popular item, 
 	like the famous movie \textit{Sherlock Holmes}, 
 	many users and entities 
 	will be involved in the U-I subgraph $|\mathcal G_{u,i|L}|$.
 \end{itemize}

Hence, in order to reduce the computation cost,
especially in the inference stage,
we should seriously take the problems (i) and (ii) into account.
In this subsection,
we customize the computation of U-I subgraphs
such that multiple items can be simultaneously evaluated 
in Section~\ref{sssec:subgraph};
and in Section~\ref{sssec:ppr},
we propose to reduce the size of subgraphs based on personalized
pagerank.

\subsubsection{Simultaneous item evaluation}
\label{sssec:subgraph}

To measure the computation cost,
we introduce the \textbf{computation graph} of 
Eq.\eqref{eq:mpnn} on the U-I subgraph
as 
\begin{equation}
	{\mathcal C}_{u,i|L} = \left(\mathcal{V}_{u,i|L}^{0:L},\mathcal{E}_{u,i|L}^{1:L}\right),
	\label{eq:newgraph}
\end{equation}
where {\small $\mathcal{V}_{u,i|L}^{0:L}=\mathcal{V}_{u,i|L}^0\cup\mathcal{V}_{u,i|L}^1\cup\cdots\cup \mathcal{V}_{u,i|L}^L$} 
is the union of nodes visited in each layer
and 
{\small $\mathcal{E}_{u,i|L}^{1:L} = \mathcal{E}_{u,i|L}^1\cup\cdots \cup\mathcal{E}_{u,i|L}^L$} 
is the union of edges.
Figure~\ref{fig:pathgraph}(a) and \ref{fig:pathgraph}(b)
graphically show the computation graphs
on the two U-I subgraphs in Figure~\ref{fig:UIgraph}.
The computation graphs indicate how messages
are propagated from user to the target item.
In addition,
the number of edges in the computation
graph can be used to indicate the computation cost.

Considering two items $i_1, i_2$ for a given user $u$,
there can be several edges appearing in both 
$\mathcal C_{u, i_1|L}$ and $\mathcal C_{u, i_2|L}$,
i.e.,
$\sum_{l=1}^L|\mathcal E_{u, i_1|L}^l\cap \mathcal E_{u, i_2|L}^l|>0$.
Take Figure~\ref{fig:pathgraph}(a)-(b) as an example, 
the two computation graphs share 5 common edges.
This property can be taken into account to save computation cost.
Specifically,
instead of directly computing on ${\mathcal C}_{u,i|L}$ in Eq.\eqref{eq:newgraph},
we propose to compute on multiple subgraphs
based on a user-centric  computation graph.
Define $\mathcal{V}_{u|L}^0\equiv \{u\}$. 
For $l=1,2\cdots L$, we recursively define
\begin{equation}
	\mathcal{E}_{u|L}^l=\big\{(n_s,r,n_o)\in \mathcal E: n_s\in \mathcal{V}_{u|L}^{l-1}\big\},
	\label{eq:uedge}
\end{equation}
as the set of triplets (edges)
whose head nodes are in $\mathcal{V}_{u|L}^{l-1}$.
And $\mathcal{V}_{u|L}^{l-1}$ is recursively defined as
\begin{equation}
	\mathcal{V}_{u|L}^l=\big\{n_o:(n_s,r,n_o)\in \mathcal{E}_{u|L}^l\big\}.
	\label{eq:unode}
\end{equation}
We provide the following proposition to show the property 
of the defined sets in Eq.\eqref{eq:uedge} and \eqref{eq:unode}.
\begin{prop}
	\begin{equation*}
		\begin{split}
			\bigcup\nolimits_{i\in\mathcal I}\mathcal{V}_{u,i|L}^l &\subset \mathcal{V}_{u|L}^l \\
			\bigcup\nolimits_{i\in\mathcal I}\mathcal{E}_{u,i|L}^l &\subset \mathcal{E}_{u|L}^l
		\end{split},
		\quad l=1,\dots, L.
		\label{eq:recursive}
	\end{equation*}
	\label{prop:union}
\end{prop}
This proposition indicates that the
nodes and edges of
individual U-I subgraphs in \eqref{eq:newgraph}
for multiple  $i\in\mathcal I$
is part of the node set $\mathcal{V}_{u|L}^l$
and edge set $\mathcal{E}_{u|L}^l$.
Since $\mathcal{V}_{u,i|L}^l$ contains the nodes
$l$-steps from $u$ and $L-l$ steps from $i$
while $\mathcal{V}_{u|L}^l$ contains all the nodes
$l$-steps from $u$,
it is easy to see that 
{\small $\mathcal{V}_{u,i|L}^l\subset \mathcal{V}_{u|L}^l, \forall i\in \mathcal I$},
and thus
{\small $\bigcup\nolimits_{i\in\mathcal I}\mathcal{V}_{u,i|L}^l \subset \mathcal{V}_{u|L}^l$}.
So do the edge sets.

Then, we transfer the computation on individual subgraphs 
into the computation on the merged sets.
Given a user $u$, we define 
\begin{equation}
	{\mathcal{C}}_{u|L}=\left(\mathcal{V}_{u|L}^{0:L},\mathcal{E}_{u|L}^{1:L}\right),
	\label{eq:usergraph}
\end{equation}
where 
{\small $\mathcal{V}_{u|L}^{0:L}=\mathcal{V}_{u|L}^0\cup\mathcal{V}_{u|L}^1\cup\cdots\cup \mathcal{V}_{u|L}^L$} 
and 
{\small $\mathcal{E}_{u|L}^{1:L} = \mathcal{E}_{u|L}^1\cup\cdots \cup\mathcal{E}_{u|L}^L$},
as an $L$-layer \textbf{user-centric computation graph}.

With Proposition~\ref{prop:union}
and the example in Figure~\ref{fig:pathgraph}(c),
we observe that 
computation graphs on U-I subgraphs 
are also subgraphs in the user-centric computation graph,
allowing us to simultaneously evaluate different items
efficiently.
Denote $\bar{D}$ as the average degree of nodes in the CKG,
the computation cost in the user-centric computation graph
Eq.\eqref{eq:usergraph}
is $O(\bar{D}^L)$.
The computation cost can be saved since
\begin{equation}
\bar{D}^L
\! \approx \! \sum\nolimits_{l=1}^L \left|\mathcal{E}_{u|L}^l\right|
\! \ll \! \sum\nolimits_{i\in\mathcal I} \sum\nolimits_{l=1}^{L}\left|\mathcal{E}_{u,i|L}^l\right| 
\! \approx \! |\mathcal I|\bar{C},
\label{eq:complexity}
\end{equation}
as will be empirically demonstrated in Figure~\ref{fig:inf}(b).

\vspace{10px}

\subsubsection{Reducing the size of subgraphs}
\label{sssec:ppr}

The size of the subgraph 
will be large when a user is connected
to some popular items.
For example,
when a user $u$ likes a famous movie 
like \textit{Sherlock Holmes},
there will be many users appear in $\mathcal V_{u|L}^2$
since \textit{Sherlock Holmes} is probably to be liked by many other users.
As $l$ increases, 
the number of nodes in $\mathcal V_{u|L}^l$ will quickly increase,
resulting in significant expense.
Therefore, it is necessary to introduce a neighborhood selection mechanism to reduce the number of nodes,
especially some irrelevant ones, 
extracted from CKG.

In principle,
we hope that 
the users, items and entities in the user-centric computation graph $\mathcal C_{u|L}$ 
is relevant to the given user $u$.
Inspired by Personalized PageRank (PPR)~\cite{page1997pagerank,haveliwala200ppr,zhang2023adaprop},
which computes the importance of nodes from a specific user's perspective through random walk,
we leverage the power of PPR
to measure the importance of different
users, items, entities to a given user.
Considering efficiency,
we use matrix multiplication
rather than random walk
to compute PPR score each nodes for each user. 
Let $\bm{M}$ be the column normalized adjacency matrix of
CKG $\mathcal G$,
that is, 
$m_{ij} = 1/D_j$, where $D_j$ the degree of node $j$ if $\exists r\in\mathcal R: (i, r, j)\in\mathcal E$,
and otherwise $m_{ij}=0$.
Denote column vector $\bm{r}_u$ be the personalized pagerank score 
of all nodes in $\mathcal V$ for the given user $u$.
We set $\bm{r}_u^{0}$ as a one-hot vector where the given user has value 1.
Then $\bm{r}_u$
can be approximately calculated by iterating
\begin{equation}
	\bm{r}_u^{k+1}=(1-\alpha)\bm{M}\times \bm{r}_u^{k}+\alpha \bm{p}_u,
	\label{eq:ppr}
\end{equation}
for a certain number of steps (e.g., 20).
Here,
$\bm{p}_u$ is a one-hot vector where the given user has value 1,
and $\alpha$ is a hyperparameter with value 0.15 indicating the restart probability.
The PPR scores $\bm{r}_u$ 
for each user,which measures which nodes are more important to the user,
can be calculated before training
as a preprocessing step.
The computation time of PPR 
is not expensive compared with the training time,
as will be displayed in Section~\ref{sssec:inference}.

After preprocessing,
the PPR score of each node on the user-centric subgraph
can be easily obtained through indexing.
Then
we can perform 
personalized neighborhood selection to select relevant and important
users, items and entities.
In order to balance the different kinds of information
like user-item interaction and different entity information,
we reduce the number of nodes
with the same head node.
Specifically,
for each head node $n_s$,
we choose top-$K$ edges $(n_s, r, n_o)$
based on the PPR score of the tail node $n_o$.
Here, $K$ is a hyperparameter 
based on the characteristics of the dataset. 
In this way, unimportant nodes are filtered out and multiple relationships can be maintained.
As a result,
we can have smaller 
user-centric computation graph and achieve higher efficiency.
The pruned user-centric computation graph is denoted as 
{\small $\widetilde{C}_{u|L} = \big( \widetilde{\mathcal{V}}_{u|L}^{0:L},\widetilde{\mathcal{E}}_{u|L}^{1:L}\big)$},
with the pruned set under the tilde.

\subsection{Algorithm and optimization}
\label{ssec:optimize}

The whole algorithm, called 
Knowledge-enhanced User-Centric subgraph Network (KUCNet),
is provided in Algorithm 1. 
We start with $\bm{h}^{0}_{u:u}=\mathbf{0}$ and $\widetilde{\mathcal{V}}_{u|L}^0=\{u\}$
in line~\ref{step:init}.
Then,
we conduct message passing in lines \ref{step:mp-start}-\ref{step:mp-end}.
In each step $l$,
we firstly 
get the edges $\bar{\mathcal{E}}_{u|L}^l$ 
connected to the sampled nodes $\widetilde{\mathcal{V}}_{u|L}^{l-1}$,
and sample the set $\widetilde{\mathcal{E}}_{u|L}^l$ from
 $\bar{\mathcal{E}}_{u|L}^l$ based on the PPR score $\bm r_u$ of user $u$ in line~\ref{step:sample}.
The node set $\widetilde{\mathcal{V}}_{u|L}^l$ is updated as the tail nodes in the sampled edges
in line~\ref{step:nodes}.
We conduct message passing for nodes in 
$\widetilde{\mathcal{V}}_{u|L}^l$ 
to aggregate the neighboring edges
$\widetilde{\mathcal{N}}_{o}^l=\{(n_s,r,n_o)\in\widetilde{\mathcal E}_{u|L}^l \}$ in lines~\ref{step:mps}-\ref{step:mpe}.
After the message passing, we 
obtain {\small $\bm{h}^{L}_{u:i}$} for different items $i\in\mathcal I$
and
use {\small $\bm{h}^{L}_{u:i}$} as the encoding 
of U-I subgraph ${\mathcal G}_{u,i|L}$.
Note that we set  {\small $\bm{h}^{L}_{u:i}=\bm 0$} if 
{\small $i\notin \widetilde{\mathcal V}_{u|L}^L$}.

\begin{algorithm}[t] 
	\caption{Knowledge-enhanced user-centric subgraph network (KUCNet)} 
	\label{alg:Framwork} 
	\begin{algorithmic}[1] 
		\REQUIRE ~~\\ 
		CKG $\mathcal G= (\mathcal V, \mathcal E)$, target user $u$, PPR score $\bm{r}_u$, layer $L$, model parameters $\mathbf{\Theta}$, hyperparameter $K$.
		\STATE initialize $\bm{h}^{0}_{u:u}=\mathbf{0}$ and $\widetilde{\mathcal{V}}_{u|L}^0=\{u\}$; \label{step:init}
		\FOR{$l=1,2\cdots L$} \label{step:mp-start}
		\STATE 
		get the edge set 
		{\small $\bar{\mathcal{E}}_{u|L}^l=\{(n_s,r,n_o)\in \mathcal E: n_s\in \widetilde{\mathcal{V}}_{u|L}^{l-1}\}$};
		\STATE
		select edges 
		{\small $\widetilde{\mathcal{E}}_{u|L}^l$}
		from {\small $\bar{\mathcal{E}}_{u|L}^l$} based on $\bm r_u$;
		\label{step:sample}
		\STATE get {\small $\widetilde{\mathcal{V}}_{u|L}^l=\{n_o:(n_s,r,n_0)\in \tilde{\mathcal{E}}_{u|L}^l\}$}; \label{step:nodes}
		\FOR{{\small $e\in \widetilde{\mathcal{V}}_{u|L}^l$} (in parallel)} \label{step:mps}
		\STATE $\bm{h}^l_{u:e}:= \delta\big(\sum\nolimits_{(n_s,r,n_o)\in \widetilde{\mathcal{N}}_{o}^l}\varphi(\bm{h}^{l-1}_{u:s},\bm{h}_r^l)\big)$ where {\small $\widetilde{\mathcal{N}}_{o}^l=\{(n_s,r,n_o)\in\widetilde{\mathcal E}_{u|L}^l \}$};
		\ENDFOR \label{step:mpe}
		\ENDFOR \label{step:mp-end}
		\RETURN $\bm{h}^L_{u:i}$ for all $i\in \mathcal I$.
	\end{algorithmic}
\end{algorithm}

For model optimization, we adopt BPR loss\cite{rendle2012bpr},
such that the given user's interacted items should get higher scores than the unobserved items:
\begin{equation}
\mathcal L =  \sum\nolimits_{(u,i,j)\in\mathcal O}  - \ln\sigma(\hat{y}_{ui} - \hat{y}_{uj}),
\label{eq:loss}
\end{equation}
where $\mathcal O=\{(u,i,j)|(u,i)\in \mathcal O^+, (u,j)\notin \mathcal O^+\}$ is the training dataset 
consisting of the observed interactions $\mathcal O^+$
with randomly generated negative item $j$.
The set of model parameters
$\mathbf{\Theta}=\big\{\{\bm{W}^l\},\{\bm{w}_\alpha^l\},\{\bm{W}_{\alpha s}^l\},\{\bm{W}_{\alpha r}^l\},\{\bm b_\alpha\} \{\bm{h}^{l}_{r}\}, l=1\dots L, r\in\mathcal R, \bm w\big\}$ in Algorithm~\ref{alg:Framwork} are randomly 
initialized and optimized by minimizing $\mathcal{L}$ in \eqref{eq:loss} with Adam stochastic gradient descent algorithm
\cite{kingma2014adam}.

\section{Experiments}

\subsection{Experimental settings}

\begin{table}[t]
	\centering
	\caption{Statistics of datasets used in this paper.}
	\vspace{-6px}
	\begin{tabular}{c|L{42px}|R{32px}|R{32px}|R{32px}|R{26px}}
		\toprule
		&       & Last-FM & Amazon-Book & Alibaba-iFashion  & DisGeNet  \\
		\midrule
		\multirow{3}{*}{UIs} & \#Users & 23,566 & 70,679 & 114,737 & 13,074 \\
		& \#Items & 48,123 & 24,915 & 30,040 & 8,947 \\
		& \#Interactions & 3,034,796 & 847,733 & 1,781,093 & 130,820 \\
		\midrule
		\multirow{3}{*}{KG} & \#Entities & 58,266 & 88,572 & 59,156 & 14,196 \\
		& \#Relations & 9     & 39    & 51 & 4 \\
		& \#Triplets & 464,567 & 2,557,746 & 279,155 & 928,517 \\
		\bottomrule
	\end{tabular}%
	\label{tab:traditional rec info}%
	\vspace{-10px}
\end{table}%

\subsubsection{Datasets}
Following \cite{wang2019kgat,wang2021learning},
we utilize three benchmark datasets for book, music, 
and fashion outfit recommendation: Last-FM, Amazon-Book and 
Alibaba-iFashion.
Last-FM contains music listening data 
collected from Last.fm online music systems, 
where tracks are viewed as items
Amazon-Book is a commonly used book dataset from Amazon-review for product recommendation.
Alibaba-iFashion is a fashion outfit dataset 
collected from Alibaba online shopping, 
where outfits are viewed as items. 
The statistics of these datasets are summarized in Table~\ref{tab:traditional rec info}.

\subsubsection{Metrics}
We employed the all-ranking strategy\cite{krichene2020sampled}.
For each user in the test set,
we calculate their preferences scores based on different models,
such as Eq.\eqref{eq:pred},
over all the items,
excluding the positive items in the training set.
We treat items that the user has not interacted with as negative items,
while items in the testing set are treated as positive.
To evaluate top-\textit{N} recommendation,
the commonly used metrics in recommendation,
recall@\textit{N} and ndcg@\textit{N},
were adopted to evaluate the results, as per \cite{krichene2020sampled}.
Recall@$N$ is defined as the ratio of the number of relevant recommended items to the total number of relevant items.
Ndcg@$N$ is defined as the sum of the discounted relevance scores of the recommended items, 
divided by the sum of the discounted relevance scores of the top-$N$ items.
Specifically,
the two metrics on a given user 
are computed by 
\begin{align}
\text{recall}@N &= \frac{{|R_{1:N}\cap T|}}{|T|}	 
\\
\text{ndcg}@N &= \frac{\sum_{i=1}^{N}\mathbb{I}(R_i\in T)\nicefrac{1}{\log_2 (i+1)}}{\sum_{i=1}^{\min(|T|,N)}\nicefrac{1}{\log_2 (i+1)}},
\end{align}
where \textit{N} is set as 20 by default, 
\textit{R} is a list of recommended items 
(excluding positive items that appear in the training set) for the given user,
\textit{T} is the test item set of the given user,
and indicator function $\mathbb{I}(x) = 1$ if $x$ is true and 0 otherwise.
We report the average metrics for all users in the test set.
For both metrics, the larger value indicates the better performance.

\subsubsection{Hyperparameters}
Hyper-parameters are selected based on
the training loss (with the maximum number
of training epochs set to 30). For KUCNet, we tune the
learning rate in $[10^{-6}, 10^{-2}]$, weight decay in $[10^{-5}, 10^{-2}]$,
dropout rate in $[0, 0.2]$, batch size in $\{10, 20, 30, 50\}$,
dimension $d$ in $\{36, 48, 64\}$, $d_{\alpha}$ for the attention weight in $\{3, 5\}$,
number of layers $L$ in $\{3, 4, 5\}$, 
number of sampling $K$ in $[20,200]$,
and the activation
function $\delta$ in $\{\text{identity}, \text{tanh}, \text{ReLU}\}$.

\subsection{Recommendation under traditional setting}
In this part,
we compare KUCNet with several baseline recommendation methods,
following the setting proposed
by \cite{wang2019kgat} and \cite{wang2021learning}.
Denote $\mathcal I_{train}$ and $\mathcal I_{test}$
as the sets of items appearing in training and testing,
respectively.
There are not new items in this setting,
meaning that all items in the testing set appeared in the training set,
i.e., $\mathcal I_{test} \subset \mathcal I_{train}$.

\subsubsection{Baselines}
\label{sssec:trans-baseline}
We compare with three types of baselines:
CF-based methods that only work on the user-item graph,
KG-based methods that 
separately learn the representations of user-item and KGs, 
and CKG-based methods that work on the collaborative KG 
with both user-item interaction and KGs.
\begin{itemize}[leftmargin=*]
	\item 
	The CF-based methods include 
	MF (matrix factorization) \cite{rendle2012bpr}, 
	FM (factorization machine) \cite{rendle2011fast} , 
	and NFM (neural FM) \cite{cao2018neural}. 
	MF factorizes the interaction network into user/item embeddings. FM also factorizes the interaction network into embeddings and captures second-order feature interactions among different embedding dimensions. 
	NFM combines the idea of factorization machines 
	with neural networks.
	\item 
	The KG-based methods included 
	RippleNet \cite{wang2018ripplenet}, 
	KGNN-LS\cite{wang2019knowledge} , 
	CKAN \cite{wang2020ckan} , 
	and KGIN \cite{wang2021learning}. 
	RippleNet  combines embedding-based and path-based methods, learning user representations by iteratively propagating representations over paths in the KG. 
	KGNN-LS   adopts GNN method to compute item embeddings on KG with label smoothing regularization of edge weights to enhance generalization.
	CKAN  separately encodes users' and items' representations by propagating entity representations in the KG with attention mechanism. 
	KGIN models user's different intents as an attentive combination of KG relations and performs the relational path-aware aggregation for both user-item graph and KG for better model capacity and interpretability.
	\item 
	The CKG-based methods included 
	CKE \cite{zhang2016collaborative}, 
	R-GCN \cite{schlichtkrull2018RGCN}, 
	and KGAT \cite{wang2019kgat}. 
	CKE is an embedding-based method with KG, which exploits semantic embeddings derived from TransR to enhance the factorization over the CKG. 
	R-GCN  is originally proposed for the KG completion task, where the user/item's representations are aggregated from the neighbors in CKG. 
	KGAT is a variant of R-GCN with an attention mechanism, allowing the importance of neighbors to be reweighted.
\end{itemize}

\subsubsection{Performance comparison}

\begin{table}[t]
	\centering
	\caption{Traditional Recommendation Performances. The best performance is shown in bold.}
	\setlength\tabcolsep{4.2pt}
	\vspace{-3px}
	\begin{tabular}{c|c|c|c|c|c|c}
		\toprule
		& \multicolumn{2}{c|}{Last-FM} & \multicolumn{2}{c|}{Amazon-Book} & \multicolumn{2}{c}{Alibaba-iFashion} \\
		& recall & ndcg  & recall & ndcg  & recall & ndcg \\
		\midrule
		MF    & 0.0724 & 0.0617 & 0.1300  & 0.0678 & 0.1095 & 0.0670 \\
		FM   & 0.0778	& 0.0644&0.1345	&0.0701	& 0.1001& 0.0602	\\
		NFM	 & 0.0829 & 0.0671	&0.1366	&0.0713	&0.1035&0.0654	\\
		\midrule
		RippleNet & 0.0791&0.0652 & 0.1336 &0.0694& 0.0960 & 0.0521 \\
		KGNN-LS & 0.0880 & 0.0642 & 0.1362 & 0.0560 & 0.1039 & 0.0557 \\
		CKAN  & 0.0812 & 0.0660 & 0.1442 & 0.0698 & 0.0970 & 0.0509 \\
		KGIN  & 0.0978 & 0.0848 & 0.1687 & 0.0915 & \textbf{0.1147} &\textbf{ 0.0716} \\
		\midrule
		CKE   & 0.0732 & 0.0630 & 0.1342 & 0.0698 & 0.1103 & 0.0676 \\
		R-GCN & 0.0743 & 0.0631 & 0.1220 & 0.0646 & 0.0860 & 0.0515 \\
		KGAT  & 0.0873 & 0.0744 & 0.1487 & 0.0799 & 0.1030 & 0.0627 \\
		\midrule
		KUCNet &\makecell[c]{\textbf{0.1205}\\{\color{black}\scriptsize$\pm$0.0002}}& \makecell[c]{\textbf{0.1078}\\{\color{black}{\scriptsize$\pm$0.0003}}} & \makecell[c]{\textbf{0.1718}\\{\color{black}{\scriptsize$\pm$0.0011}}} & \makecell[c]{\textbf{0.0967}\\{\color{black}{\scriptsize$\pm$0.0008}}} & \makecell[c]{0.1031\\{\color{black}{\scriptsize$\pm$0.0005}}} & \makecell[c]{0.0663\\{\color{black}{\scriptsize$\pm$0.0006}}} \\
		\bottomrule
	\end{tabular}%
	\label{tab:tradition result}%
	\vspace{-7px}
\end{table}%

Based on the experimental results reported in Table~\ref{tab:tradition result},
the following observations can be made:

KUCNet outperforms all other baselines in terms of recall@20 and ndcg@20 on the Last-FM and Amazon-Book datasets.
This suggests that the U-I subgraph, which integrates collaborative signals and item knowledge, 
is a powerful and effective tool for personalized recommendation.
Additionally, KUCNet can effectively learn the complex structure of subgraphs, 
exploring the potential interests of users 
from known interactions.
 
 However, on the Alibaba-iFashion dataset, 
 KUCNet did not perform better than the baseline 
 and performed poorly like most GNN methods. 
 The reason for this is that the characteristics of the dataset are different from those of Last-FM and Amazon-Book. 
 The KG of Last-FM and Amazon-Book offer denser and richer information than that of Alibaba-iFashion, 
 which makes it more helpful in capturing the essential connections between items. 
 Furthermore, in Alibaba-iFashion, 
 the first-order connectivity (fashion outfit, including, fashion staff) dominates the KG triplets, 
 making it difficult to discover relationships between items and explore user interests on subgraphs. 
 Therefore, simple collaborative filtering and embedding-based methods become more effective.

Overall, using KG/CKG performs better than traditional CF-based methods, 
indicating the importance of side information in KG. 
GNN-based methods, such as KGIN, KGAT, KGNN-LS, and CKAN, perform better than embedding-based and path-based methods, 
and also significantly better than R-GCN. 
Since R-GCN is not originally designed for recommendation, 
it fails to model user-item relationship properly.


\subsection{Recommendation with new items}
\label{ssec:inductive}

In addition to the traditional recommendation scenario,
we set up a special recommendation scenario to analyze
the recommendation capability of methods on the new items.
Specifically,
we evaluate the ability to recommend new items
that appear in the KG but have not interacted with any users.

To construct the dataset,
we randomly select one-fifth of the items from all items to form $\mathcal I_{test}$, 
whose interaction history is removed from the training set, 
and the other four-fifths of items form $\mathcal I_{train}$. 
Therefore, 
$\mathcal I_{test} \cap \mathcal I_{train} = \emptyset$.
The user-item graph was then split into training and testing data based on whether the item appeared in $\mathcal I_{train}$ and $\mathcal I_{test}$.
Note that even in the test phase, 
we do not know the interaction history of these items in the testing set. 
Hence, the models could only find and recommend these items through the CKG known during training. 
This scenario allows us to evaluate the performance of the methods in recommending new items.

\subsubsection{Baselines}
In addition to the baseline methods described in Section~\ref{sssec:trans-baseline},
which are not very powerful in dealing with new items,
we introduce new baselines in this section
that are more capable of handling new items.
In particular,
these new baselines do not learn specific embeddings
for users and items.

The first new baseline is PPR (personalized PageRank) \cite{haveliwala200ppr}, 
which computes personalized PageRank scores of nodes in the CKG for each user. 
We can then conduct recommendation directly based on the PPR scores of users to items.

The second one is PathSim \cite{sun2011pathsim}, 
which pre-defines some meta-paths for each dataset 
and extracts the meta-paths between users and items from the CKG as features. 
The recommendation is then predicted based on the extracted features.

The third is RED-GNN \cite{zhang2022knowledge}, 
a GNN-based method designed for the KG completion task
with subgraph learning.
By leveraging the subgraphs, RED-GNN
does not need to learn embeddings,
and thus can work on new items.

%
%

\subsubsection{Performance comparison}

\begin{table}[t]
	\centering
	\caption{Comparison of different methods for Recommendation on new items.
		The best performance is shown in bold numbers.}
	\vspace{-4px}
	\setlength\tabcolsep{4.3pt}
	\label{tab:inductive result}
	\begin{tabular}{c|c|c|c|c|c|c}
		\toprule
		& \multicolumn{2}{c|}{Last-FM} & \multicolumn{2}{c|}{Amazon-Book} & \multicolumn{2}{c}{Alibaba-iFashion} \\
		         & recall & ndcg & recall & ndcg & recall & ndcg \\
		\midrule
		MF & 0     & 0     & 0     & 0     & 0     & 0 \\
		FM & 0.0012 & 0.0007 & 0.0026 & 0.0010 & 0 & 0  \\
		NFM &  0.0125 & 0.0068  & 0.0006 &0.0003 &0 &0 \\
		\midrule

		RippleNet & 0.0005 & 0.0004 & 0.0011 & 0.0005 & 0.0007 & 0.0004  \\
		KGNN-LS &  0     & 0      & 0.0001     &0.0001 &  0.0001  &0.0001   \\
		CKAN & 0.0005 & 0.0005 & 0.0005  & 0.0003   &0.0003 &0.0002  \\
		KGIN & {0.2472} & {0.2292} & {0.0868} & {0.0446} & 0.0010 & 0.0004 \\
		\midrule
		
		CKE & 0     & 0     & 0     & 0     & 0   & 0 \\
		R-GCN & 0.0616 & 0.0372 & 0.0001 & 0.0001 & 0.0001 & 0.0001 \\
		KGAT & 0 & 0 & 0.0001     & 0.0001     & 0 & 0 \\
		
		\midrule
		PPR &  0.2274  &  0.1919  & 0.0301 & 0.0167 & 0.0001 & 0.0001 \\
		PathSim &0.5248 & 0.5308  & 0.2053 & 0.1491 & 0.0202 & 0.0088 \\
		REDGNN  & 0.5284 & 0.5425 & 0.2187 & 0.1633 & 0.0072 & 0.0043  \\
		
		\midrule 
		KUCNet & \makecell[c]{\textbf{0.5375}\\{\color{black}{\scriptsize$\pm$0.0010}}} & \makecell[c]{\textbf{0.5573}\\{\color{black}{\scriptsize$\pm$0.0012}}} & \makecell[c]{\textbf{0.2237}\\{\color{black}{\scriptsize$\pm$0.0020}}} &\makecell[c]{\textbf{0.1685}\\{\color{black}{\scriptsize$\pm$0.0009}}} & \makecell[c]{\textbf{0.0269}\\{\color{black}{\scriptsize$\pm$0.0005}}} & \makecell[c]{\textbf{0.0149}\\{\color{black}{\scriptsize$\pm$0.0003}}} \\
		\bottomrule
	\end{tabular}%
	\vspace{-7px}
\end{table}%

In this setting,
where the new items for testing have no interaction with the users in the training data, 
the experimental results  in Table~\ref{tab:inductive result} show that traditional CF methods, 
such as MF, FM, and NFM, 
and KG/CKG-based methods using node embedding,
such as RippleNet, KGNN-LS, CKAN, CKE,  and KGAT
perform poorly on all datasets. 
This indicates that these methods are not effective on the new items.
KGIN, which combines CF and KG embedding, performs significantly better than other baselines on the Last-FM and Amazon-Book datasets but has poor performance on the Alibaba-iFashion dataset.

The baselines introduced in these part are significantly better than
those in Section~\ref{sssec:trans-baseline}.
PPR and PathSim perform better than traditional methods but worse than RED-GNN on Last-FM and Amazon-Book datasets.
KUCNet outperforms all other baselines on all datasets, indicating that the U-I subgraph, 
which integrates collaborative signals and item knowledge, 
is a powerful and effective tool for personalized recommendation even for the new items. 
Additionally, KUCNet can effectively learn the complex structure of subgraphs, 
exploring the potential interests of users from known interactions.

It is interesting to note that almost all methods fail on the Alibaba-iFashion dataset, except for KUCNet and PathSim. 
This indicates that the KG information for this dataset is not rich 
and substantial enough to reveal the essential relationship between items.

%

\subsection{Recommendation with new items and new users}

 In the previous experiments, 
 the datasets used do not include side-information about users,
 which limits our ability to conduct experiments involving new users. 
In this part,
 we conduct a new set of experiments that simulates the scenario of encountering new users. 
 We choose the task of disease gene prediction, 
 which can be regarded as a recommendation problem where diseases are users and genes are items,
 \textcolor{black}{to show the generalization of KUCNet across different domains}.
The curated disease-gene associations from the DisGeNet database\cite{pinero2016disgenet}
 and a biological KG \cite{wang2023KDGene} centered on diseases and genes
 are used to build the dataset. 
  The KG involves multiple relations, 
  including disease-gene, disease-disease, gene-gene, gene-GO, and gene-pathway from well-known biomedical databases. 
  The disease-gene relation can be seen as user-item interaction, 
  while disease-disease represents the connection between different users, 
  similar to the social networks in reality.
  The statistics of this dataset is summarized in the rightmost column of Table~\ref{tab:traditional rec info}.

\subsubsection{Settings}
Similarly,  we set up two new settings:
\begin{itemize}[leftmargin=*]
	\item new item (gene):
	same as recommendation setting in Section~\ref{ssec:inductive}, 
	we randomly split the set of items into five parts.
	Then one fifth of the items 
	are put into $\mathcal I_{test}$,
	while the other four-fifths in $\mathcal I_{train}$.
	The 5 different splits are regarded as 5-fold validation to evaluate the models.
	This setting connects new type of genes to given diseases.
	
	\item new user (disease):
	we also randomly split the set of users into five parts.
	In each fold, 
	the one-fifth of new users have 
	no interaction history on the training set
	and we need to recommend items to these new users.
	 This setting aims to connect genes to new types of diseases.
\end{itemize}

Through these experiments, 
we gain valuable insights into how our model 
can utilize side-information on the user side
to provide accurate recommendations for new users. 

\begin{table}[t]
	\centering
	\caption{Disease Gene prediction performances. The best performance is shown in bold.}
	\vspace{-5px}
	\renewcommand\arraystretch{0.95}
	\begin{tabular}{C{38px}|C{37px}|C{37px}|C{37px}|C{37px}}
		\toprule
		  & \multicolumn{2}{c|}{new item} & \multicolumn{2}{c}{new user} \\
		  & recall & ndcg  & recall & ndcg \\
		\midrule
		 MF    & 0.0000 & 0.0000 & 0.0123 & 0.0086   \\
		FM    & 0.0007    & 0.0003   & 0.0238  & 0.0165   \\
		 NFM   & 0.0038  & 0.0033  & 0.0296   & 0.0211  \\
	    \midrule
		  RippleNet & 0.0023    & 0.0011   & 0.0027   & 0.0018  \\
		 KGNN-LS & 0.0017    & 0.0006   & 0.0080   & 0.0048  \\
		  CKAN  & 0.0189   & 0.0086 & 0.0244  & 0.0138  \\
		  KGIN  & 0.0989    & 0.0568  & 0.0031  & 0.0023  \\
		\midrule
		 CKE   & 0.0001   & 0.0000  & 0.0072  & 0.0066  \\
		 KGAT  & 0.0032   & 0.0015   & 0.0364    & 0.0264    \\
		 R-GCN & 0.0598    & 0.0294   & 0.1498   & 0.1014   \\ 
		 \midrule
	      PPR   & 0.1293   & 0.0665   & 0.0194  & 0.0156   \\
	      PathSim & 0.2023     & 0.1506    & 0.2810   & 0.2144   \\
		  REDGNN & 0.2341    & 0.1523    & 0.2821  & 0.2154   \\
		  \midrule
		 KUCNet & \makecell[c]{\textbf{0.2574}\\{\color{black}$\pm$0.0121}} & \makecell[c]{\textbf{0.1791}\\{\color{black}$\pm$0.0101}} & \makecell[c]{\textbf{0.2883}\\{\color{black}$\pm$0.0025}} & \makecell[c]{\textbf{0.2274}\\{\color{black}$\pm$0.0015}}\\
		\bottomrule	
\end{tabular}%
\label{tab:Disease result}%
\vspace{-10px}
\end{table}%

\subsubsection{Performance comparison}

As shown in Table~\ref{tab:Disease result}, in new item setting, 
traditional recommendation methods perform poorly,
similar to the new item recommendation scenario.
PPR, PathSim and RED-GNN are better at exploring newly emerged items.
Compared with them, 
our method achieved significant advantages by
the specially designed U-I subgraphs for recommendation.

In the new user setting, 
effective utilization of user-KG can achieve good results due to the lack of historical records for new users. 
Among GNN-based methods, 
KGAT performed better since it takes users into the KG to build a  CKG 
and can leverage similarity between users. 
R-GCN and RED-GNN also utilize the connections between users, achieving better results. 
However, our method still has obvious advantages, 
proving that we can effectively utilize user-side information to provide more accurate recommendations for new users.

\subsection{Efficiency analysis}

In this section,
we analyze the efficiency of our proposed KUCNet in terms
of three aspects:
the learning curve,
the number of parameters,
and the inference time.

\subsubsection{Learning curves}
\label{ssec:learning curve}

\begin{figure}[t]
	\centering
	\subfigure[Curves of recall@20.]
	{\includegraphics[width=0.493\linewidth]{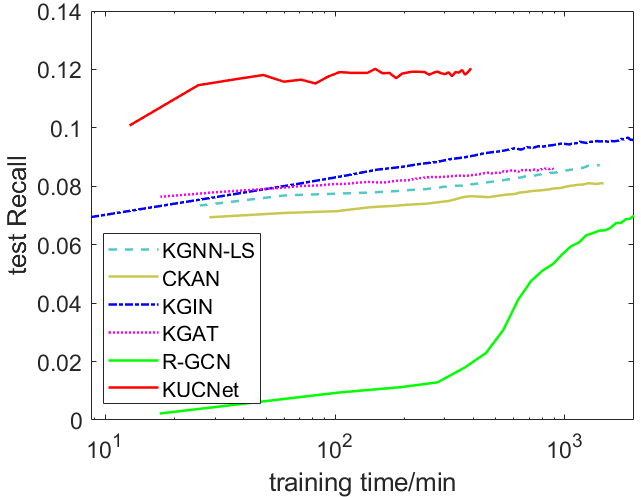}}
	\hfill
	\subfigure[Curves of ndcg@20.]
	{\includegraphics[width=0.493\linewidth]{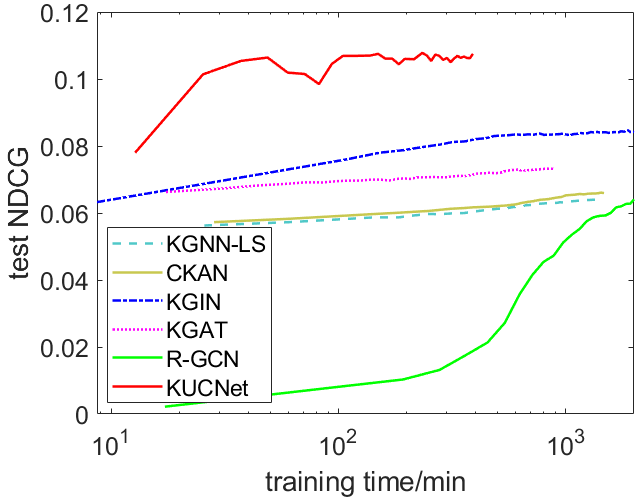}}
	\vspace{-6px}
	\caption{Learning curves on Last-FM.
	The two metrics Recall@20 and ndcg@20 have similar trends for the different methods.}
	\label{fig:training time}
\end{figure}

Figure~\ref{fig:training time} shows the learning curves on Last-FM,
demonstrating that
KUCNet can achieve better performance in a shorter period of training time compared to other methods that use GNN.
In addition, other methods require longer time for convergence
to obtain the best embedding representation.
Among these, R-GCN requires the longest time
since it is not specifically designed for user-item interaction prediction
and needs extra time to learn other relationships.

\subsubsection{Model parameters}

\begin{figure}[t]
	\centering
	\includegraphics[width=0.95\linewidth]{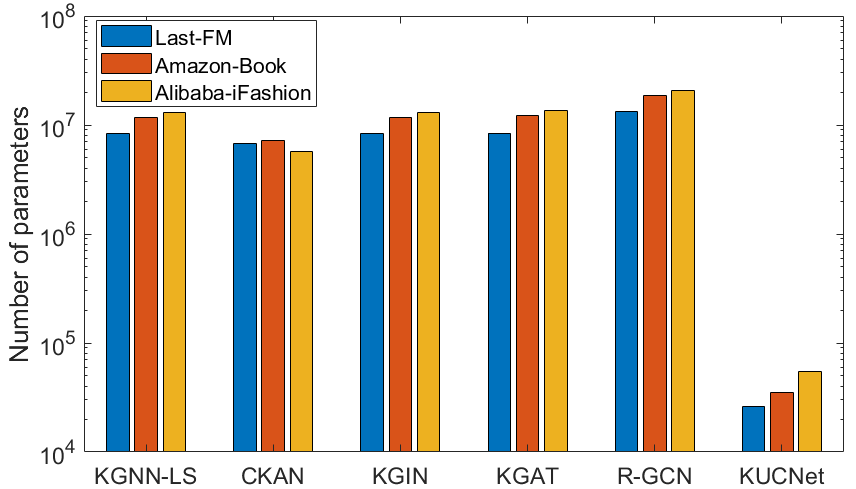}
	\vspace{-5px}
	\caption{Comparison of model parameters on three datasets.}
	\label{fig:params}
	\vspace{-7px}
\end{figure}

Figure~\ref{fig:params} shows
the numbers of model parameters obtained on three datasets.
KUCNet has significantly fewer model
parameters than the other methods using KG.
The key reason is that
KUCNet does not require learning specific node embeddings,
while other methods all have embeddings for each node.
Such a small number of parameters also helps our model 
achieve the best results within a shorter training time, 
as already shown in Figure~\ref{fig:training time}.

\subsubsection{Improvement on inference time}
\label{sssec:inference}

\begin{figure}[t]
	\centering
	\subfigure[Inference time.]
	{\includegraphics[width=0.49\linewidth]{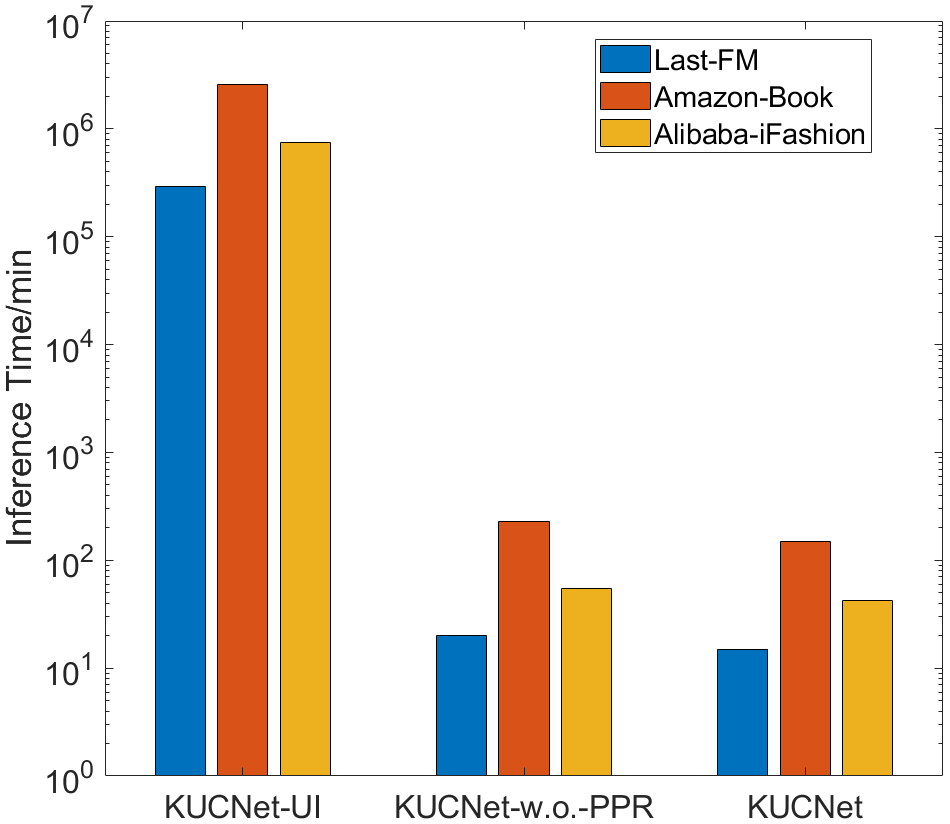}}
	\subfigure[Number of edges.]
	{\includegraphics[width=0.49\linewidth]{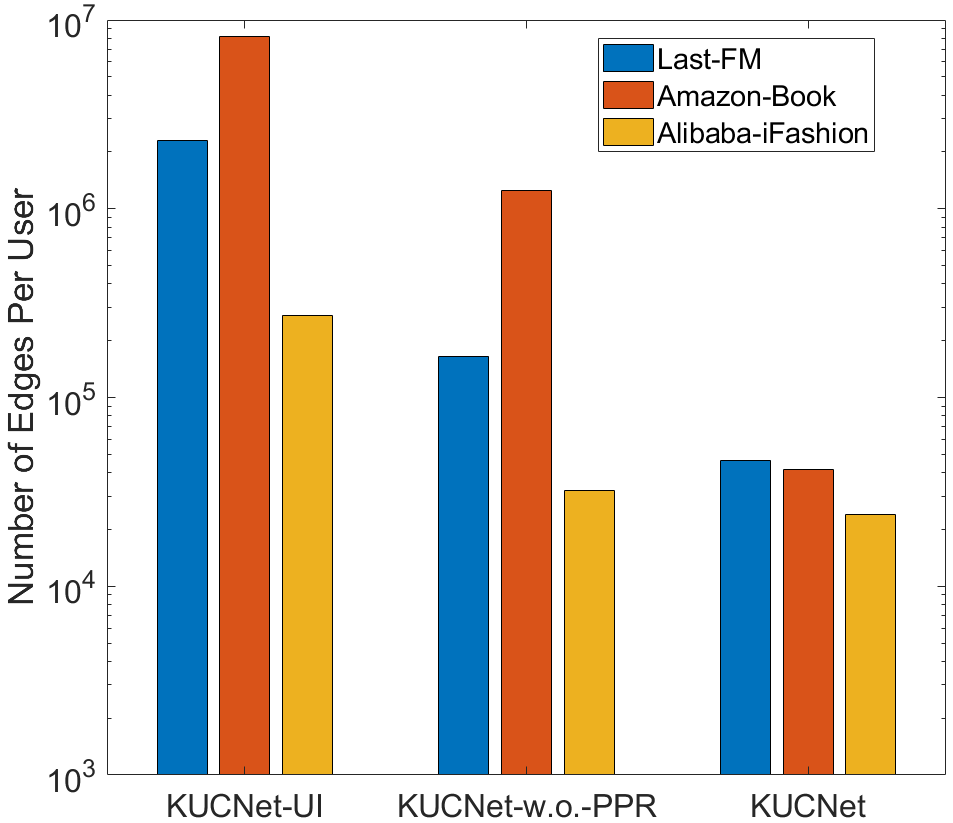}}
	\vspace{-10px}
	\caption{Inference time and number of edges on different computation graphs.}
	\label{fig:inf}
\end{figure}

In Section~ \ref{sssec:subgraph},
we discussed that the inference time of computing on each U-I subgraph through Eq.\eqref{eq:mpnn} is very expensive,
and we proposed two approaches to address the challenges of
multiple evaluation items and subgraph size, respectively.
In this part,
we compare the inference time 
and the number of edges in the computation graphs
of these implementations
in Figure~\ref{fig:inf}.
We observe that the inference time is dramatically saved on the user-centric computation graph (KUCNet-w.o.-PPR)
compared with directly computing individual U-I subgraphs (KUCNet-UI).
Furthermore, with PPR sampling, 
the inference time of the model (KUCNet) 
can be further improved. 
The direct solution, 
KUCNet-UI, on individual U-I graphs has millions of edges to compute on per user. 
In comparison, the user-centric computation graph in KUCNet-w.o.-PPR has much fewer numbers of edges than individual U-I graphs, demonstrating Eq.\eqref{eq:complexity}. 
With PPR sampling, 
KUCNet can further reduce the size of computation graphs and thus reduce the inference cost.

Moreover, we compare the running time of PPR computing process with training and inference time in Table~\ref{tab:pprtime}. 
Since the runtime is basically proportional to the number of users, 
less than fifty minutes were spent on the dataset Alibaba-iFashion with the most users, 
which is significantly lower than the training time required to achieve optimal performance. 
As a one-time preprocessing, the overhead is completely acceptable.

\begin{table}[t]
	\centering
	\vspace{-5px}
	\caption{Running time comparison of preprocessing on PPR values, training and inference stages in minutes.}
	\vspace{-6px}
	\begin{tabular}{c|c|c|c}
		\toprule
		& {Last-FM} & {Amazon-Book} & Alibaba-iFashion \\
		\midrule
		PPR & 8  &  25   & 46  \\
		Training &204 &335&  304\\
		Inference & 15 &150 &  42\\
		\bottomrule
	\end{tabular}%
	\label{tab:pprtime}%
	\vspace{-10px}
\end{table}%

\subsection{Interpretability}
\label{ssec:exp:interpret}

\begin{figure}[ht]
	\centering
	\color{black}
	\subfigure[Example on Last-FM.]
	{\includegraphics[width=0.85\linewidth]{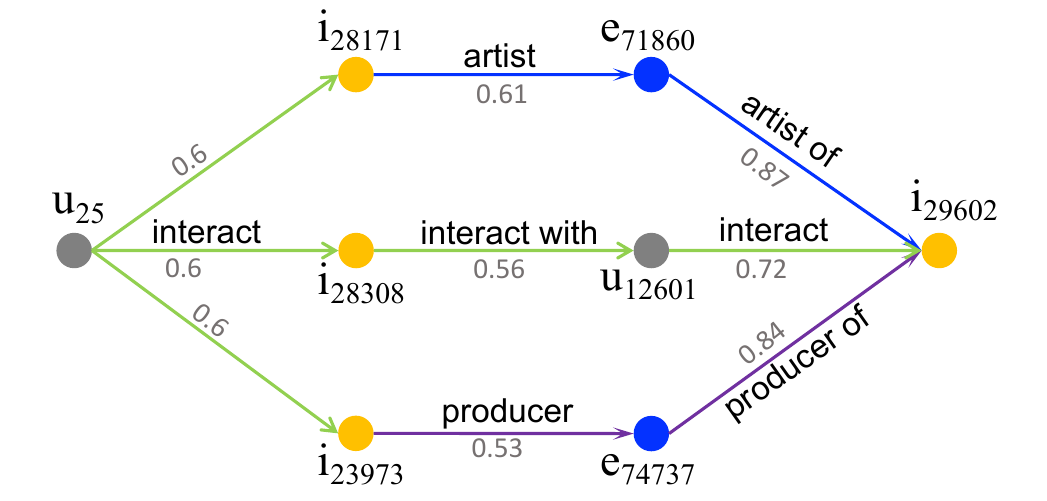}}
	
	\subfigure[Example on new-Last-FM.]
	{\includegraphics[width=0.85\linewidth]{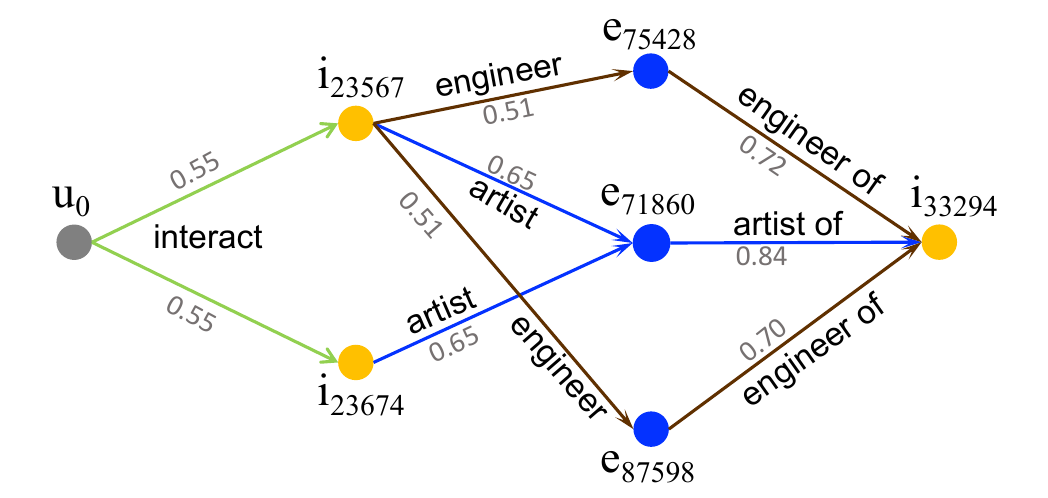}}
	
	\subfigure[Example on DisGeNet with new item (gene).]
	{\includegraphics[width=0.85\linewidth]{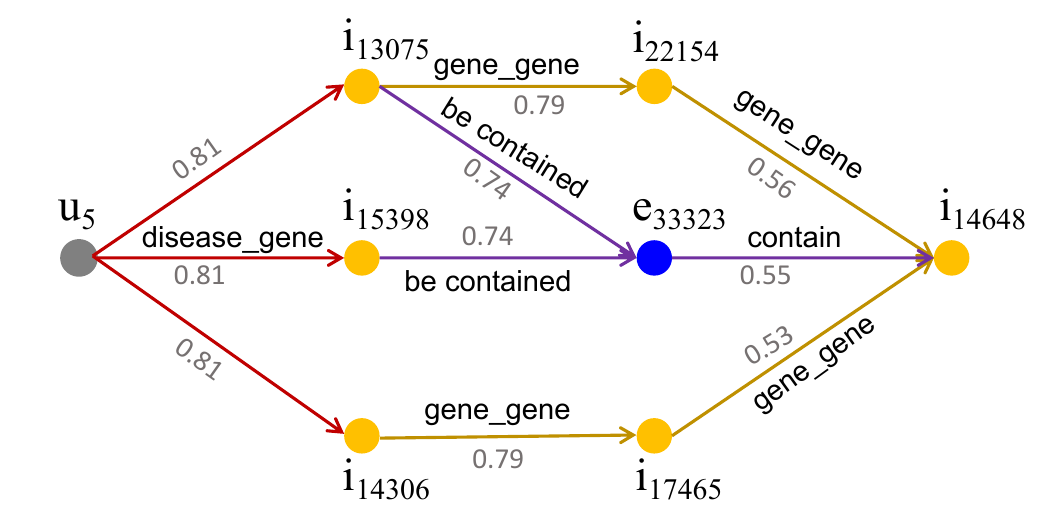}}

	\subfigure[Example on DisGeNet with new user (disease).]
	{\includegraphics[width=0.85\linewidth]{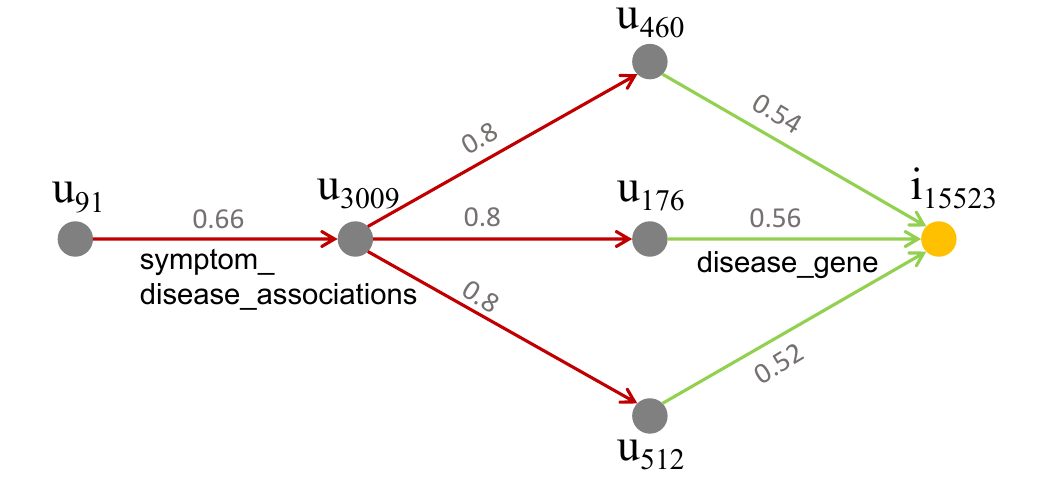}}
	\caption{Visualization of the learned U-I subgraphs by KUCNet.}
	\label{fig:vis}
	\vspace{-6px}
\end{figure}

In this section, 
we discuss the interpretability of KUCNet
and visualize some exemplar learned U-I subgraphs 
by KUCNet with \textit{L}=3  on the Last-FM 
and DisGeNet datasets in Figure \ref{fig:vis}.
These subgraphs are extracted from the pruned user-centric
computation graph,
and edges with attention weights lower than 0.5 are eliminated.
{\color{black}These attention weights allow us to pinpoint the triples that play a crucial role 
in the recommendation, as indicated by the gray numbers in the Figure.}
We show four instances in different settings as follows:
\begin{itemize}[leftmargin=*]
	\item Figure \ref{fig:vis}(a) shows a case 
	in traditional recommendation scenario, 
	explaining why we recommend item 29602 to user 25 on Last-FM.
	As shown, 
	user 25 has interacted with three important items with IDs 28171, 28308, 23973, respectively.
	They either share the same artist/producer with  item 29602
	or have both been interacted with the same user. 
	So item 29602 is chosen to be recommended to the user.
	
	\item Figure \ref{fig:vis}(b) shows a recommendation instance in the setting recommending new items.
	Since new items have not interacted with any user, 
	KUCNet only explores the information on KG,
	automatically eliminating interference 
	caused by collaborative signals during training.
	
	\item Figure \ref{fig:vis}(c) shows an example in new item (gene) setting.
	Since it is unknown which diseases are directly related to the gene 14648,
	the model employs the interactions and shared inclusion relations between genes to identify it.
	
	\item Figure \ref{fig:vis}(d) shows an example in new user (disease) setting. Propagating through 
	two hops on the subgraph, 
	a new disease 91 finds other three diseases, 
	that have similar symptom,
	and they all connect to gene 15523,
	indicating that this gene may be closely related 
	to the disease.
	
\end{itemize}

By visualizing the learned subgraphs, 
we can 
find the most relevant information supporting the recommendation,
interpret how KUCNet captures the relationships between users and items, 
and explain how it utilizes the KG to make accurate recommendations
in different recommendation scenarios.

\subsection{Ablation studies}

In this section, 
we discuss several results of ablation studies 
of KUCNet.
Recall@20 is used to measure the performance.

\subsubsection{Influence of K}

Table \ref{tab:effct of K} shows the effect of 
the number of sampling $K$ for KUCNet
in both traditional and new item recommendation scenarios
(In the new item scenario, the dataset is prefixed with ``new-").
 The experimental results indicate that moderate sampling can significantly improve recommendation performance.
If $K$ is too small, only a small amount of information will be utilized,
but 
when $K$ is too large,
too many edges will be sampled, 
introducing noisy information and bringing higher costs. 
The optimal number of sampling varies with different datasets.

\begin{table}[t]
	\centering
	\caption{Comparison of different sampling number $K$.}
	\label{tab:effct of K}
	\vspace{-6px}
	\renewcommand\arraystretch{1.3}
	\begin{tabular}{C{63px}|c|c|c|c|c}
		\hline
		K       & 20     & 30     & \textbf{35}     & 40     & 50     \\ \hline
		Last-FM & 0.1200 & 0.1202 & \textbf{0.1205} & 0.1199 & 0.1198 \\ \hline
	\end{tabular}
	
	\vspace{4px}
	\begin{tabular}{C{63px}|c|c|c|c|c}
		\hline
		K           & 100    & 110    & \textbf{120}    & 130    & 140    \\ \hline
		Amazon-Book & 0.1702 & 0.1707 & \textbf{0.1718} & 0.1714 & 0.1703 \\ \hline
	\end{tabular}
	
	\vspace{4px}
	\begin{tabular}{C{63px}|c|c|c|c|c}
		\hline
		K         & 30     & 40     & \textbf{50}     & 60     & 70     \\ \hline
		new-Last-FM & 0.5339 & 0.5368 & \textbf{0.5375} & 0.5369 & 0.5362 \\ \hline
	\end{tabular}
	
	\vspace{4px}
	\begin{tabular}{C{63px}|c|c|c|c|c}
		\hline
		K         & 150    & 160    & \textbf{170}    & 180    & 190 \\ \hline
		new-Amazon-Book & 0.2175 & 0.2197 & \textbf{0.2237} & 0.2196 & 0.2172   \\ \hline
	\end{tabular}
\end{table}

\subsubsection{Influence of L}

In this part, we analyze the influence of model depth $L$
for KUCNet in the recommendation scenario.  
When $L < 3$, 
there will be
no length $L$ paths 
connecting given user and majority candidate items.
However, when $L$ is too large, the size of the user-centric graph may lead to high memory and time cost. 
Hence,
we show the results when $L\in\{3,4,5\}$ .

\begin{table}[t]
	\centering
	\vspace{-4px}
	\caption{Comparison of different model depth $L$.}
	\vspace{-6px}
	\begin{tabular}{c|c|c|c}
		\toprule
		L & {3} & {4} & 5 \\
		\midrule
		Last-FM & \textbf{0.1205} & 0.1125 &0.1150    \\
		Amazon-Book & \textbf{0.1718} & 0.1667   & 0.1688   \\
		Alibaba-iFashion & \textbf{0.1031} & 0.1004  &0.1015   \\
		\midrule
		new-Last-FM & \textbf{0.5375}  & 0.5216   & 0.5331   \\
		new-Amazon-Book & \textbf{0.2237}  &0.1952   & 0.2030   \\
		new-Alibaba-iFashion & 0.0057 & 0.0056  &\textbf{0.0269}   \\
		\bottomrule
	\end{tabular}%
	\label{tab:effect of L}%
	\vspace{-7px}
\end{table}%

We show the performances on three recommendation datasets 
to analyze the influence of $L$ in Table \ref{tab:effect of L}.
The results show that
when $L=3$, the performances on Last-FM 
and Amazon-Book achieve the best, 
since almost all candidates items are covered and
the subgraph already has complex structures 
and rich semantic information.
However, on Alibaba-iFashion, especially in new item setting, the three-layer model can not 
provide good recommendations, 
so it is necessary to deepen the model to five layers to 
explore more candidate items and higher-order semantic information.

\subsubsection{Effect of PPR}
In this experiment, we study a variant (named KUCNet-random) 
that uses random sampling instead of PPR score 
to study the effectiveness of PPR sampling.
Experiments are performed on the two datasets in
two different settings  in Table \ref{tab:random}.
The PPR sampling method achieves better results than the random sampling method, 
which proves the effectiveness of PPR sampling.
Hence,
selecting relevant edges to by PPR score to the target user
can bring benefit to the subgraph learning problem.

\subsubsection{Influence of Attention}

We also study a variant (named KUCNet-w.o.-Attn) 
that does not use the attention mechanism in \eqref{eq:message func}.
Table~\ref{tab:random} shows that,
without the attention mechanism, 
the performances decrease,
verifying the importance of using an
attention mechanism to control the importance of edges.

\begin{table}[t]
	\centering
	\caption{Comparison of different variants of KUCNet.}
	\renewcommand\arraystretch{0.95}
	\vspace{-6px}
	\begin{tabular}{C{70px}|C{40px}|C{40px}|C{40px}}
		\toprule
		method & KUCNet-random & KUCNet-w.o.-Attn & KUCNet \\
		\midrule
		Last-FM & 0.1181  & 0.1193& \textbf{0.1205}     \\
		Amazon-Book & 0.1655  & 0.1672 &   \textbf{0.1718 }  \\
		\midrule
		new-Last-FM &0.5293 & 0.5348 & \textbf{0.5375}\\
		new-Amazon-Book &0.2142 &0.2172 & \textbf{0.2237} \\
		\bottomrule
	\end{tabular}%
	\label{tab:random}%
	\vspace{-7px}
\end{table}%

These ablation studies provide insights into the importance of different components in KUCNet and offer guidance for optimizing the model.

\section{Conclusion}

In this paper, we proposed KUCNet, 
a novel knowledge-enhanced personalized recommendation method 
that constructs a U-I subgraph for each user-item pair, 
and then adopts a user-centric subgraph network with an attention-based GNN
and pruned by PPR.
Experimental results demonstrate that KUCNet outperforms state-of-the-art 
KG-based and CF-based recommendation methods, 
while maintaining high efficiency and interpretability. 
Specifically, KUCNet achieves significant improvements in recommendation scenario 
for new items. The proposed method can effectively capture the most relevant information 
for each user-item pair, while avoiding the noise and redundancy in the KG.

Our proposed method has several advantages over existing methods, 
including leveraging both user-item interactions and relevant KG information
of most importance, 
adopting a user-centric subgraph network for efficient and personalized recommendation, 
and achieving accurate and interpretable recommendations. 
KUCNet is an effective and efficient method for knowledge-enhanced recommendation, 
which can be applied to various recommendation scenarios and has the potential 
to enhance the user experience in various domains,
e.g., drug-drug interaction prediction \cite{zhang2023emerging}.
In the future, we can
consider the combination of KUCNet with Large Language Models
to enable more effective recommendation on text-rich items 
like news articles, blog posts, health treatment, etc.

\section*{Acknowledgment}

This project was supported by the National Natural Science Foundation of China (No. 92270106).

\clearpage

\bibliographystyle{IEEEtran}
\bibliography{bib}

\end{document}